\documentclass[11pt]{article}
\usepackage{float}
\usepackage{tabularx}
\usepackage{amsmath,amssymb}
\usepackage{graphicx}
\usepackage{authblk}
\usepackage[margin=1in,letterpaper]{geometry}
\usepackage{authblk}
\usepackage{blindtext}
\usepackage{array}
\usepackage[table]{xcolor}
\usepackage{multirow}
\usepackage{caption}
\usepackage{subcaption}
\usepackage{makecell}
\usepackage{hyperref}
\hypersetup{colorlinks,bookmarksopen,bookmarksnumbered,
linkcolor=cyan,urlcolor=black,citecolor=red}

\usepackage[numbers,sort&compress]{natbib}

\graphicspath{{figs/}}

\graphicspath{{figs/}}

\begin{document}

\title{Detecting Beyond the Standard Model Interactions of Solar Neutrinos in Low-Threshold Dark Matter Detectors}
\author{Thomas Schwemberger\thanks{tschwem2@uoregon.edu}}
\author{Tien-Tien Yu\thanks{tientien@uoregon.edu}}
\affil{\normalsize\it Department of Physics and Institute for Fundamental Science, University of Oregon, Eugene, Oregon 97403, USA}
\maketitle

\begin{abstract}
As low-threshold dark matter detectors advance in development, they will become sensitive to recoils from solar neutrinos which opens up the possibility to explore neutrino properties. We predict the enhancement of the event rate of solar neutrino scattering from Beyond the Standard Model interactions in low-threshold DM detectors, with a focus on silicon, germanium, gallium arsenide, xenon, and argon-based detectors. We consider a set of general neutrino interactions, which fall into five categories: the neutrino magnetic moment as well as interactions mediated by four types of mediators (scalar, pseudoscalar, vector, and axial vector), and consider coupling these mediators to either quarks or electrons. Using these predictions, we place constraints on the mass and couplings of each mediator and the neutrino magnetic moment from current low-threshold detectors like SENSEI, Edelweiss, and SuperCDMS, as well as projections relevant for future experiments such as DAMIC-M, Oscura, Darwin, and ARGO. We find that such low-threshold detectors can improve current constraints by up to two orders of magnitude for vector mediators and one order of magnitude for scalar mediators.
\end{abstract}

{
  \hypersetup{linkcolor=black}
  \tableofcontents
}

\section{Introduction}\label{sec:intro}
The Weakly Interacting Massive Particle (WIMP) paradigm has driven the landscape of dark matter (DM) research on both the experimental and theoretical fronts for the last several decades.
The tightening of constraints in the WIMP parameter space has motivated the development of low-threshold dark matter (DM) detectors targeted at DM masses below those predicted by the canonical WIMP models, which sit at the ${\cal O}(10-100 {\rm eV})$ mass scales. These low-threshold DM detectors are sensitive ${\cal O}(\rm eV)$ energy depositions, opening up access to sub-GeV DM candidates. 
The DM detectors optimized for WIMP searches rely on the detection of DM-nuclear scattering, in which the recoiling nucleus imparts some detectable signature in the form of heat, light, or ionization. It has been long understood that neutrinos, specifically solar neutrinos, can mimic these same signatures through the coherent scattering of neutrinos on nuclei; this has been historically called the ``neutrino floor" or ``neutrino background" (see {\it e.g.}~\cite{Billard:2013qya} and references therein). More recently, it was demonstrated that low-threshold detectors which observe ionization, {\it e.g.} electron, signatures are also subject to this neutrino background~\cite{essig_solar_2018}. Although the background is due to neutrino-nuclear scattering, a fraction of the nuclear recoil energy can be converted into electron recoil energy, thus leading to an ionization signal. 

The 2017 observation of coherent elastic neutrino-nucleus scattering (CEvNS) by COHERENT~\cite{akimov_observation_2017} motivates the discussion of the potential to unearth new Beyond the Standard Model (BSM) physics in the neutrino sector. 
This has led to a series of efforts to look for generalized BSM interactions of neutrinos with the SM in a variety of detectors. Previous work has focused on detectors with thresholds above ${\cal O}(100~\rm{eV})$~\cite{denton2018plan, Majumdar:2021vdw} or nucleon only interactions~\cite{Li:2022jfl}.
In this work, we make a systematic study of BSM neutrino interactions in low-threshold DM detectors, down to the bandgap energies of ${\cal O}(\rm{eV})$, using a simplified models framework to encode the neutrino non-standard interactions (NSIs).\footnote{Note that we are using the terminology ``NSI" for the simplified-models framework rather than the effective field theory style framework typically used to quantify new physics in the neutrino sector (see {\it e.g.}~\cite{dev_neutrino_2019, AristizabalSierra:2018eqm,Bischer:2019ttk}). Thus, neutrino magnetic moments fall under our definition of NSI.} 
The advantage to using low-threshold DM detectors to study NSI is two-fold. Low-threshold ionization experiments, as discussed above, are sensitive to both electron and nuclear recoils, the latter as a result of the conversion of nuclear recoil energy into electron recoil energy. This makes these detectors sensitive to a wider range of models, some of which can evade the astrophysical constraints that rely primarily on electron interactions.
We will calculate the sensitivity to both quark and electron couplings independently as well as a neutrino magnetic moment.
Furthermore, there is an additional enhancement of the scattering rate which is inversely proportional to the nuclear recoil energy from which low-threshold detectors, with their sensitivity to small energy depositions, profit tremendously. We also calculate the sensitivities in noble liquid detectors using xenon and argon where the much larger exposures can overcome the relative inefficiency of converting nuclear recoil energy into an ionization signal.  

The difficulty in detecting CEvNS and low-energy neutrino-electron scattering lies in running an experiment with both low detector noise and a low energy threshold. Recent technological advancements in semiconductor targets such as the use of Skipper CCDs~\cite{Tiffenberg:2017aac} or high bias potential voltages~\cite{Romani:2017iwi,EDELWEISS:2019vjv} have led to sensors with sub-electron noise and energy thresholds down to the band gap, ${\cal O}({\rm eV})$, of the semiconductor. Existing experiments to date have provided early data with an exposure of $\sim 10$ g-days~\cite{sensei_collaboration_sensei_2020,edelweiss_collaboration_first_2020,supercdms_collaboration_constraints_2020} which can place constraints on light vector boson-mediated models such as $U(1)_{B-L}$ comparable to those of Borexino~\cite{harnik_exploring_2012, galbiati_new_2008}. Similarly, noble liquid detectors such as XENON1T also set competitive constraints on such models~\cite{Boehm_xenon}. These constraints will only strengthen with the next-generation of low-threshold detectors which will have increased exposure.  
With larger Skipper CCD experiments planned, it will be important to understand the signals they may be sensitive to both from neutrinos and dark matter.

The rest of this paper is organized as follows. In Sec.~\ref{sec:scattering} we present the expressions for the neutrino-nuclear and neutrino-electron scattering rates in both the SM and five NSI simplified models. In Sec.~\ref{sec:detectors}, we discuss the signatures of the scattering events in semiconductor (Si, Ge, and GaAs) and noble liquid (Xe and Ar) materials used in low-threshold detectors. Here, we include the discussion of ionization efficiencies and charge yields in each material. In Sec.~\ref{sec:results} we discuss the effects of the simplified models on the event rates calculated in Sec.~\ref{sec:detectors} and present our projected sensitivities to the NSI simplified model parameters. We discuss the interpretation of our results in relation to complete ultraviolet models in Sec.~\ref{sec:discussion} and conclude in Sec.~\ref{sec:conclusions}. 
\section{Neutrino Scattering Rates}\label{sec:scattering}
\begin{figure}[h]
    \centering
    \includegraphics[width=0.8\textwidth]{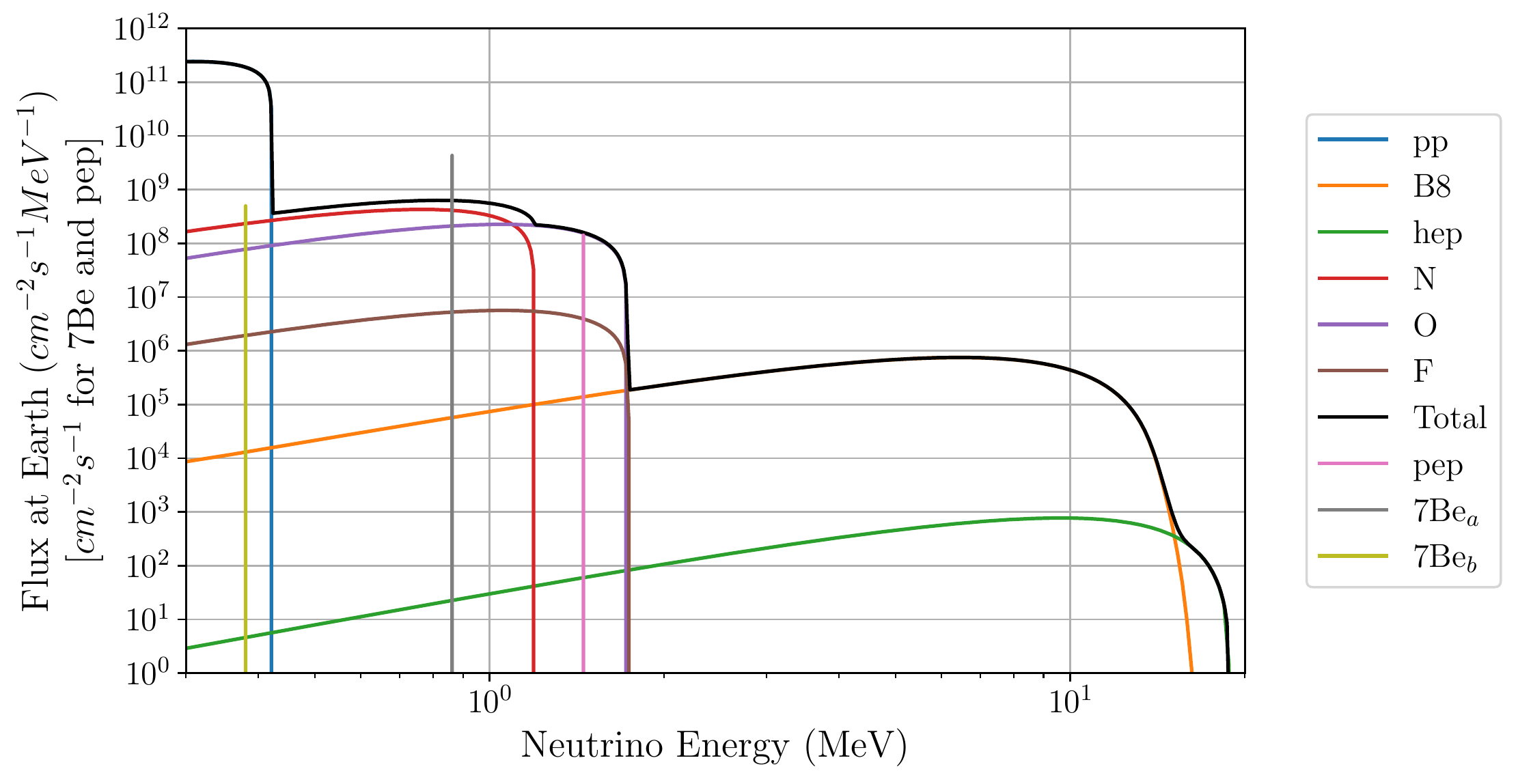}
    \caption{Differential neutrino flux components from the {\it pp} and {\it CNO} cycles using the high-metallicity solar neutrino model BS05(OP)~\cite{bahcall_new_2005, bahcall_how_2005}. The black line does not include the monochromatic components which are included individually in the calculations which follow.}
    \label{fig:solarnuflux}
\end{figure}

Neutrinos are produced in the sun via a series of nuclear processes, which are broadly categorized as {\it pp} or {\it CNO}, depending on the elements involved. These nuclear processes constitute the different components of the solar neutrino flux, as shown in Fig.~\ref{fig:solarnuflux}.
To determine the measured event rate from solar neutrino scattering in a detector, we follow the procedure outlined in \cite{essig_solar_2018}, and sum the contributions from each component of the solar flux, themselves determined by integrating over their energy spectra. We use the fluxes given by the high-metallicity solar neutrino model BS05(OP) \cite{bahcall_new_2005, bahcall_how_2005}. Specifically, in order to produce a (kinetic) recoil energy $E_R$, the incident neutrino must have at least the energy given in Eq.~\eqref{E_min} where $m_T$ as the target mass (either electron or nucleus) and we have explicitly kept the neutrino mass ($m_\nu$) to leave open the possibility of heavy sterile neutrinos. 
The minimum neutrino energy is therefore given by
\begin{equation} \label{E_min}
    E_\nu^{min} = \frac{1}{2}\left(E_R + \sqrt{E_R^2+2 E_R m_T \left(\frac{m_\nu^2}{m_T^2}+1 \right) + 4m_\nu^2}\right)\, .
\end{equation}
We integrate the differential neutrino flux, $\frac{dN_\nu}{dE_\nu}$, from this minimum energy to the highest energy neutrino in each solar component weighting by the differential cross-section for either nuclear or electron recoils and multiply by the number of targets per kg of detector material ($N_T$) to get the recoil rate in events per kg-yr,
\begin{equation} \label{integral}
    \frac{dR}{dE_{R}}=N_T\int_{E_\nu^{min}}\frac{d\sigma}{dE_{R}}\frac{dN_\nu}{dE_\nu}dE_\nu\, ,
\end{equation}
where $\frac{d\sigma}{dE_{R}}$ is the differential cross-section of neutrino scattering. In what follows, we present the expressions for this differential cross-section in both the SM and in five representative NSI simplified models. 
\subsection{The Standard Model}

The SM differential cross-sections for nuclear and electron recoils are

\begin{equation} \label{SM_cross_N}
    \frac{d\sigma}{dE_{N}}=\frac{G_F^2}{4\pi}Q_v^2m_N\left(1-\frac{m_N E_{N}}{2E_\nu^2}\right)F^2(E_{N})
\end{equation}
\begin{equation} \label{SM_cross_e}
    \frac{d\sigma}{dE_{e}}=\frac{G_F^2m_e}{2\pi E_\nu^2}\left[ 4s_w^2(2E_\nu^2 + E_e^2 - E_e(2E_\nu + m_e)) \pm 2s_w^2(E_e m_e - 2E_\nu^2) + E_\nu^2 \right]\, ,
\end{equation}
where $G_F$ is the Fermi constant, $s_w$ is the sine of the Weinberg angle, $Q_v=N-Z(1-4s_w^2)$ is the weak nuclear hypercharge (for $N$ neutrons and $Z$ protons), and the $\pm$ distinguishes electron neutrinos ($-$) from the other flavors ($+$). $F(E_N)$ is the Helm nuclear for factor as defined in \cite{lewin_review_1996}, which is approximately unity at the energy scales we consider. We leave the discussion of our treatment of neutrino oscillation to Sec.~\ref{NSI_rates}.

\subsection{Non-Standard Neutrino Interactions} \label{NSI_rates}
Low-threshold detectors like SENSEI, Edelweiss and SuperCDMS, open up access to the lower range of recoil energies for which neutrino scattering produces a detectable signature. Therefore, we focus on a set of simplified infrared models for NSI which modify the differential cross-section at low energies.
Arguably the simplest way to introduce a NSI is to give the neutrino a magnetic moment, $\mu_\nu$, via the interaction
\begin{equation} \label{L_mu}
    \mathcal{L} \supset \mu_\nu \bar{\nu}\sigma^{\alpha\beta}\partial_\beta A_\alpha\nu\, .
\end{equation}
As of 2022, the strongest constraint on the magnetic moment of $\mu_\nu < 2.8 \times 10^{-11}\mu_b$, where $\mu_b=\sqrt{4\pi\alpha}/2m_e$ is the Bohr magneton, is from Borexino~\cite{the_borexino_collaboration_limiting_2017}. Note the inclusion of this interaction provides an enhancement to both the CEvNS and electron recoil rates, though as we will see in Sec.~\ref{sec:detectors} the former is suppressed by both the yield function and nuclear mass. 

The enhancement of the neutrino-nucleus cross-section is given by

\begin{equation} \label{sig_mu}
    \frac{d\sigma_\mu}{dE_R} = \mu_\nu^2 \alpha Z^2 F^2(E_R) \left(\frac{1}{E_R} -\frac{1}{E_\nu}\right)
\end{equation}
where $F(E_R)$ is, again, the Helm nuclear form factor as detailed in \cite{lewin_review_1996}. As before, $F(E_R)\sim 1$ for the recoil energies of interest. The neutrino electron cross-section is similar to Eq.~\ref{sig_mu} but without the nuclear charge number ($Z$) or form factor.

For full generality, we consider set of simplified models in which the couplings of the mediator to neutrinos and the target (nucleus, electron) along with the mass of the mediator are free parameters.  This approach permits us to remain agnostic about the underlying theory and allows the reader the map our results onto a wide-range of UV-models. We calculate the enhancement (or suppression) to the scattering cross-section resulting from NSI mediated by scalar, pseudoscalar, vector, or axial vector propagators, as shown in Eq.~\eqref{L_sim}. 
For scalars and pseudoscalars, we include line 1 of Eq.~\eqref{L_sim} while for vectors and axial vectors we use line 2. Line 3 denotes the operators $\mathcal{O}_i$ coupling the field ($i$) to fermions with the appropriate Lorentz structure as shown in Table~\ref{tab:cross_sections}.
\begin{equation} \label{L_sim}
\begin{split}
\mathcal{L} \supset (&g_{\nu P} \phi \Bar{\nu}_R\nu_L + h.c.)+(g_{\nu S} \phi \Bar{\nu}_R\nu_L + h.c.)\\
&+g_{\nu V} \Bar{\nu}_L \gamma^\mu \nu_L Z'_\mu
+g_{\nu A} \Bar{\nu}_L \gamma^\mu \nu_L Z'_\mu \\
&+g_{ei} \mathcal{O}_i(\Bar{e}, e) + g_{qi} \mathcal{O}_i(\Bar{q}, q)
\end{split}
\end{equation}

\begin{table}[ht]
    \centering
    \begin{tabular}{ c c c}
     \hline \hline
     Mediator & Couplings & $\mathcal{O}_i(\Bar{f}, f)$\\
     \hline
     Scalar (S)      & $g_{\nu S}, g_{e S}, g_{q S}$ & $\phi \Bar{f} f$ \\  
     Pseudoscalar (P) & $g_{\nu P}, g_{e P}, g_{q P}$ & $-i\gamma^5\phi \Bar{f} f$ \\  
     Vector   (V)    & $g_{\nu V}, g_{e V}, g_{q V}$ & $Z'_\mu \Bar{f} \gamma^\mu f$ \\
     Axial Vector (A) & $g_{\nu A}, g_{e A}, g_{q A}$ & $-Z'_\mu \Bar{f} \gamma^\mu \gamma^5 f$ \\
     \hline \hline
    \end{tabular}
    \caption{Couplings, and operators for the four $(S,P,V,A)$ generalized mediator models. Note the relative minus-sign for the CP-odd interactions.}
    \label{tab:cross_sections}
\end{table}
 The electron recoil cross sections from the NSI contribution without the SM are given in Eq.~\eqref{eq:e_cross}. As in \cite{cerdeno_physics_2016}, we neglect terms which are higher order in $E_R/E_\nu$ since we focus on signals in low threshold detectors where $E_R/E_\nu \lesssim 10^{-2}$. The subscripts $(S, P, V, A)$ correspond to the scalar, pseudoscalar, vector, and axial vector mediated interactions. We consider NSIs where the neutrino coupling is generation independent, but the generalized case simply considers the recoil rate from each neutrino flavor weighted by their oscillation probability.

\begin{equation} \label{eq:e_cross}
\begin{split}
    \left.\frac{d\sigma_e}{dE_R}\right\vert_{S} \hspace{0.5mm} &= \frac{g_{\nu S}^2 g_{e S}^2 E_R m_e^2}{4\pi E_\nu^2(2E_Rm_e + m_S^2)^2} \\
    \left.\frac{d\sigma_e}{dE_R}\right\vert_{P} &= \frac{g_{\nu P}^2 g_{e P}^2 E_R^2 m_e}{8\pi E_\nu^2(2E_Rm_e + m_P^2)^2} \\
    \left.\frac{d\sigma_e}{dE_R}\right\vert_{V} &= \frac{\sqrt{2}G_Fm_e g_v g_{\nu V}g_{e V}}{\pi (2E_Rm_e + m_V^2)} + \frac{m_e g_{\nu V}^2 g_{e V}^2 }{2\pi(2E_Rm_e + m_V^2)^2}\\
    \left.\frac{d\sigma_e}{dE_R}\right\vert_{A} &= \frac{\sqrt{2}G_F m_e g_a g_{\nu A}g_{e A}}{\pi (2E_Rm_e + m_A^2)} + \frac{m_e g_{\nu A}^2 g_{e A}^2 }{2\pi(2E_Rm_e + m_A^2)^2}
\end{split}\, ,
\end{equation}
where $g_v$ and $g_a$ are the weak vector and axial couplings, and $m_{e,S,P,V,A}$ are the masses of the electron, scalar, pseudoscalar, vector, and axial vector respectively. Eq. \eqref{eq:e_cross} does not include the effects of electron binding energy which can have a measurable effect at low energies~\cite{CHEN2017656}. We will revisit this effect in Sec.~\ref{sec:results} where its effect on sensitivities is relevant. Note however that they are coupling solar neutrinos to electrons, so we must consider a weighted average over the same-flavor and mixed-flavor couplings: $g_{v,e}=2s_w^2+1/2$, $g_{a,e}=+1/2$ and $g_{v,\mu \tau}=2s_w^2-1/2$, $g_{a,\mu \tau}=-1/2$. We do this with the typical 2-flavor neutrino mixing approximation in Eq.~\eqref{eq:mixing} where $\sin^2(\theta_{12})=0.31$ and $\Delta m_{12}^2=7.5 \times 10^{-5}{\rm eV}^2$ are the mixing angle and mass splitting respectively. $L$ is the Earth-Sun distance in km and $L_{osc} = 2.48{~\rm km} \times E_\nu / \Delta m_{12}^2$ with $E_\nu$ in GeV

\begin{equation} \label{eq:mixing}
    P(\nu_e \rightarrow \nu_\mu) = \sin^2(2\theta_{12})\sin^2(\pi L/L_{osc})
\end{equation}
For most solar neutrinos, $L \gg L_{osc}$ and $\sin^2(L/L_{osc}) \rightarrow \left<\sin^2(L/L_{osc}) \right> = 1/2$ is a good approximation when we integrate over the neutrino energy spectrum. However, this is not true for monochromatic solar neutrino channels such as $^7$Be and pep sources since we are effectively integrating a delta function and are sensitive to the exact value of $\Delta m_{12}^2$. Therefore, we use the full expression, Eq.~\ref{eq:mixing}, in our calculations.

The nuclear recoil cross-sections are given by

\begin{equation} \label{eq:n_cross}
\begin{split}
    \left.\frac{d\sigma_N}{dE_R}\right\vert_{S} \hspace{0.5mm}  &= \frac{F^2(E_R) {Q'}_s^2 E_R m_N^2}{4\pi E_\nu^2(2E_Rm_N + m_S^2)^2} \\
    \left.\frac{d\sigma_N}{dE_R}\right\vert_{P} &= 0 \\
    \left.\frac{d\sigma_N}{dE_R}\right\vert_{V} &= -\frac{F^2(E_R) G_Fm_NQ_vQ'_v(2E_\nu^2-E_Rm_N)}{2\sqrt{2}\pi E_\nu^2 (2E_Rm_N + m_V^2)} + \frac{F^2(E_R) m_N{Q'}_v^2(2E_\nu^2-E_Rm_N)}{4\pi E_\nu^2(2E_Rm_N + m_V^2)^2}\\
    \left.\frac{d\sigma_N}{dE_R}\right\vert_{A} &= \frac{F^2(E_R) G_Fm_NQ'_a[Q_a(2E_\nu^2+E_Rm_N)-Q_vE_RE_\nu]}{2\sqrt{2}\pi E_\nu^2(2E_Rm_N + m_A^2)} + \frac{F^2(E_R) m_N {Q'}_a^2 (2E_\nu^2+E_Rm_N)}{4\pi E_\nu^2(2E_Rm_N + m_A^2)^2}
\end{split}\, ,
\end{equation}
where the effective nuclear couplings $Q'_{S,V,A}$ and $Q_{v, a}$ used in the CEvNS enhancement are listed in Eq.~\eqref{eq:coherence} and $S_N$ is the nuclear spin weighted over the isotopes present in the detector.

\begin{equation} \label{eq:coherence}
\begin{split}
    {Q'}_s &= g_{\nu S}g_{q S}(14A+1.1Z) \\
    {Q'}_v &= g_{\nu V}g_{q V}(3A) \\
    {Q'}_a &= g_{\nu A}g_{q A}(0.3S_N) \\
    {Q }_v &= A-(2-4s_w^2)Z \\
    {Q }_a &= 1.3S_N
\end{split}\, .
\end{equation}
In the above, $A, Z$ correspond to the atomic mass and number, respectively.

\begin{figure}[h]
    \centering
    \includegraphics[width=0.6\textwidth]{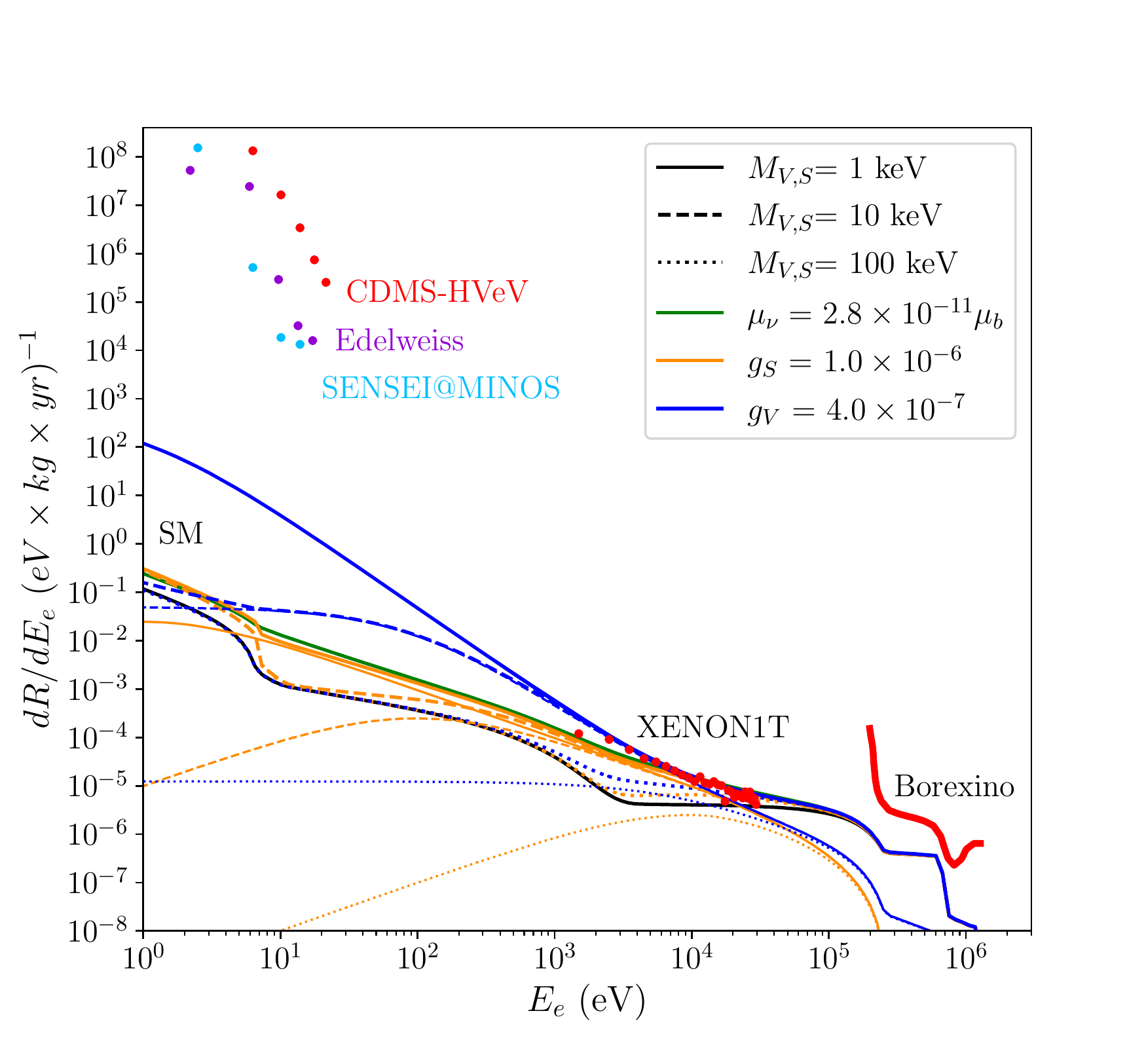}
    \caption{Combined electron and nuclear recoil rates in silicon from vector (blue) and scalar (gold) mediators and a neutrino magnetic moment (green). Nuclear rates are shifted to the rates at the equivalent electron energy to account for the yield function as discussed in Sec.~\ref{sec:detectors}. We plot the NSI electron recoil rates with thin lines to illustrate the limiting behavior at low-energies. The black line is the SM. Borexino data is from \cite{PhysRevLett.107.141302} and XENON1T from \cite{xenon_collaboration_excess_2020}, while SENSEI@MINOS, Edelweiss, and CDMS-HVeV are from \cite{sensei_collaboration_sensei_2020,edelweiss_collaboration_first_2020,supercdms_collaboration_constraints_2020} respectively. The couplings are chosen to approximately saturate the XENON1T constraints.}
    \label{fig:recoil_rates}
\end{figure}

The strengths of low-threshold detectors relative to traditional WIMP direct detection experiments is at low energies. To this end, we note the limiting behavior of each of these models, but also note that at such low energies the SM background is dominated by CEvNS which enhances both the signal and background. In the case of the neutrino magnetic moment, the recoil rate increases as $1/E_R$ without bound, though the neutrinos which produce such soft recoils also tend to be less energetic, so the $-1/E_\nu$ term reduces this gain. The scalar rate also scales as $1/E_R$ until $E_R\lesssim m_\phi^2/m_T$ where $m_T$ is the target mass, below which the electron recoil rate falls proportional to $E_R$, however, below $\sim 100$ eV nuclear recoils dominate and the rates regain their power-law form. The pseudoscalar rate does not enjoy the same low energy enhancement so the detectors we discuss here will be less sensitive to it. Both vector and axial vector mediators produce a differential rate which scales as $1/E_R^2$ and plateaus at $E_R\sim m_\phi^2/m_T$ below which nuclear recoils dominate again. This makes vector and axial vector models particularly interesting for low threshold detectors. We plot the recoil rates for the magnetic moment as well as 1, 10, and 100 keV scalar and vector mediators in Fig.~\ref{fig:recoil_rates} to illustrate these behaviors.

\section{Ionization Signals in Low Threshold Detectors} 
\label{sec:detectors}
Once a neutrino has scattered off of a nucleus or electron, the recoil energy of the target must be translated into a detectable ionization signal. In this section, we summarize that process for both electron and nuclear recoils in semiconductors and noble liquid detectors. For scattering off of electrons, one must first translate the theoretical quantity of electron recoil energy into the experimentally measured number of electrons (or photoelectrons, for the case of noble liquid detectors). The exact procedure depends on the detector, which we will describe below.

For nuclear recoil, there is an additional step of converting the nuclear recoil energy into electron recoil energy. 
To determine the nuclear recoil rate, we consider the energy a nuclear recoil $(E_R)$ must have to produce the same effect as an electron recoil with energy $E_e$. We use the convention $E_e = Y(E_R)\times E_R$ where $Y(E_R)$ is the yield function which depends on nuclear recoil energy and the material of the detector. 
The electron ionization signal from a nuclear recoil is then calculated by binning the differential electron energies corresponding to a differential nuclear recoil element. This is calculated by differentiating the electron energy as a function of nuclear recoil energy.
\begin{equation}
    \frac{dE_e}{dE_R} = Y(E_R) + E_R \frac{dY(E_R)}{dE_R}
\end{equation}
Then by a change of variables
\begin{equation} \label{diff_rate}
    \frac{dR_e}{dE_e} = \frac{dR_N}{dE_R} \times \frac{1}{Y(E_R) + E_R \frac{dY(E_R)}{dE_R}}
\end{equation}
where $R_e$ and $R_N$ are the effective electron ionization rate and nuclear scattering rates respectively.
The rate in Eq.~\eqref{diff_rate} is then binned like the direct electron recoils, noting that the right-hand side is a function of $E_R$ so the bin boundaries must be adjusted to match using the inverse of the yield function. 

At high energies, the quenching factor is well-approximated by the Lindhard model~\cite{lindhard1963integral}, 
\begin{equation} \label{eq:lindhard}
\begin{split}
    Y_{Lindhard}(E_R) &= \frac{k g(\epsilon)}{1 + k g(\epsilon)} \\
    g(\epsilon)  &= 3\epsilon^{0.15} + 0.7 \epsilon^{0.6} + \epsilon \\
    \epsilon     &= 11.5 Z^{-7/3}E_R
\end{split}\, ,
\end{equation}
where $Z$ is the atomic number of the recoiling nucleus and $E_R$ is in keV. Although the original description set $k=0.133 Z^{2/3}A^{-1/2}$, with $A$ as the mass number of the nucleus, experimental data point to a range of values for $k$. Therefore, $k$ is typically treated as a free-parameter. 

However, the Lindhard model breaks down at the energies of interest to this work. 
Combined with the lack of experimental data at the lowest energies, we must extrapolate the yield function to a range of cut-off energies in an attempt to span all possible low-energy limits.
To do this we define three yield functions (high, fiducial, low) which we set to vanish at three respective cut-off energies, $E_c$, and smoothly match them to the high-energy model with an analytic extrapolation. Where possible, we use more modern corrections to the Lindhard model which include the effects of binding energy to more closely match low-energy data from Refs.~\cite{chavarria_measurement_2016, scholz_measurement_2016, collar_germanium_2021}. In the following sections we discuss these extrapolations and updates to the Lindhard model at low energies. Note that the sensitivity of the low threshold detectors is dependent both on the magnitude and slope of the yield function as seen in Eq.~\eqref{diff_rate}, and therefore has a strong dependence on the choice of extrapolation. We examine this dependence in detail in Sec.~\ref{sec:yield_dep} and Appendix~\ref{app:yield}. 

\subsection{Semiconductors} 
\label{sec:sc_yield_functions}
In low-threshold semiconductor experiments, such as those using Skipper CCDs, the ionization signal comes in the form of quantized electrons, $n_e$. The electron recoil energy, $E_e$, can be translated into quanta by binning the electron energies starting from the band gap energy with a bin width equal to the average electron-hole pair production energy. In other words, for a recoil with energy $E_e$, $n_e=1+\lfloor{(E_e-E_{\rm gap})/\varepsilon}\rfloor$, where $\lfloor x\rfloor$ rounds $x$ down to the nearest integer, $E_{\rm gap}$ is the bandgap energy of the material, and $\varepsilon$ is the mean energy per electron-hole pair. This is a fairly good approximation, accurate to about $\sim 10\%$~\cite{Essig:2015cda} of more sophisticated models (see {\it e.g.} Refs.~\cite{Ramanathan:2020fwm,Rodrigues:2020xpt} and references therein). The relevant energies are summarized for the three semiconductors of interest in Table~\ref{tab:semi_energies}. The procedure to convert nuclear recoil energy into quantized electrons is more complicated. First, the nuclear recoil energy is translated to electron energy with a yield efficiency (sometimes called quenching factor) which carries a large theoretical uncertainty and has not been measured at energies below several hundred eV. 

\begin{figure}[ht]
    \centering
    \includegraphics[width=\textwidth]{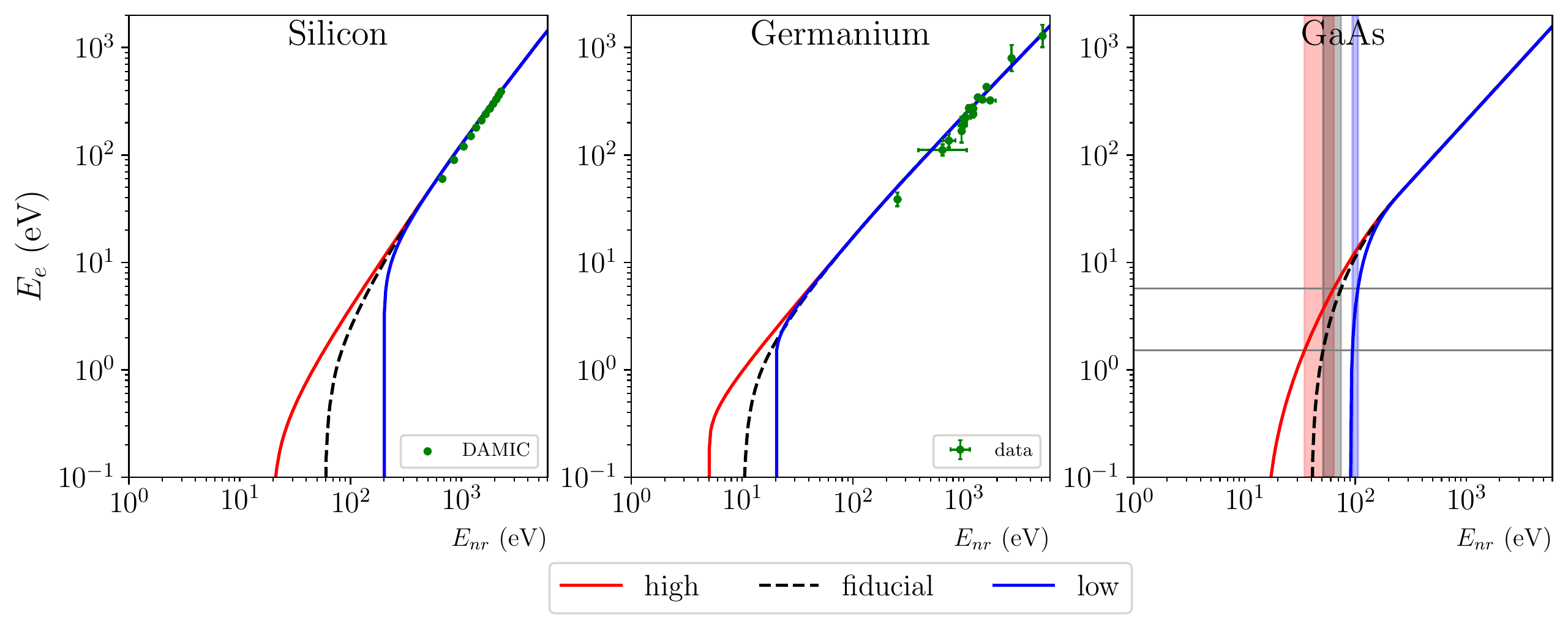}
    \caption{Yield functions in silicon, germanium, and gallium arsenide. Data points for silicon are from the DAMIC collaboration~\cite{chavarria_measurement_2016} while germanium is from~\cite{scholz_measurement_2016, collar_germanium_2021}. The vertical bands in the right frame illustrate the sensitivity of the recoil rate to the slope of the yield function. The wider bands imply that a larger range of recoil energies correspond to an event which produces the same number of electrons.}
    \label{fig:yield}
\end{figure}

The Lindhard model makes several approximations which begin to break down at the energies of interest, as discussed in Ref.~\cite{sarkis_study_2020}. For example, the Lindhard model neglects the binding energy of the electrons, assumes that the energy transferred to the electrons is significantly smaller than that transferred to the recoiling ion, and that the nuclear kinetic energy is small.  Ref.~\cite{sarkis_study_2020} revisits the calculation of Lindhard in silicon and germanium using a first-principles approach and relaxing the above approximations. Note that this work assumes a constant binding energy. Near the transition region, there exists empirical measurements by DAMIC~\cite{chavarria_measurement_2016} for silicon and Ref.~\cite{scholz_measurement_2016, collar_germanium_2021} for germanium. We use an ionization model which combines the first-principles approach of Ref.~\cite{sarkis_study_2020} and the empirical measurements. For illustrative purposes, we adopt as our fiducial models an updated calculation from Ref.~\cite{SarkisMagCEvNS2021,Sarkis:2021agv}, where they allow for the binding energy to vary. They also use an enhanced model of the electronic stopping power which takes into account Coulomb repulsion effects as well as high-energy effects from the Bohr stripping criterion. We extend this model to different cut-off energies to illustrate the sensitivity of the detectors to small changes in the yield function. The functions are plotted in Fig.~\ref{fig:yield}. Our extrapolations are fit using a power law: $Y(E_R) = a(x-E_c)^b$ where $E_c$ is the cut-off energy and $a$ and $b$ are chosen to match the high energy model at $600$ eV in silicon and $100$ eV in germanium (roughly 10 times the cut-off in the model developed by Ref.~\cite{SarkisMagCEvNS2021,Sarkis:2021agv}). In Si(Ge), our deviations extend the cut-off to 20(5) eV for the high-yield case and 200(20) eV in the conservative case.

In the absence analogous first-principles calculations for gallium arsenide, we extrapolate the Lindhard model, Eq.~\eqref{eq:lindhard}, below 254 eV to a choice of three cut-off energies (15, 40, and 90 eV), but use a different model for the extrapolation since the low energy behavior is smoother than the models of Ref.~\cite{SarkisMagCEvNS2021,Sarkis:2021agv}. Here we choose an exponential of the from $Y(E_R)=a(1-e^{(E_c-E_R)/b})$ where $a$ and $b$ are again chosen to match the high energy model at the branching energy (254 eV). The lines in Fig.~\ref{fig:yield} are the average of the Ga and As yield function and our results are reported as a sum of the contributions from each component, but we keep the yield functions independent up to the final event rate calculations.

\begin{table}[h]
    \centering
    \begin{minipage}{0.45\textwidth}
    \centering
    \begin{tabular}{ c c c }
     \hline \hline
     Material & $E_g$ [eV] & $\varepsilon$ [eV] \\
     \hline
     Si~\cite{vavilov1962radiation}   & 1.2  & 3.8 \\  
     Ge~\cite{vavilov1962radiation}   & 0.7  & 3.0 \\  
     GaAs~\cite{cusano1964radiative}  & 1.52 & 4.2 \\
     \hline \hline
    \end{tabular}
    \end{minipage}
    \begin{minipage}{0.45\textwidth}
    \centering
    \begin{tabular}{ c c c c }
     \hline \hline
     Material & $W$ [eV] & $f_R$ & $f_e$ \\
     \hline
     Xe~\cite{PhysRevD.92.032005}  & 13.8  & 0 & 0.83 \\  
     Ar~\cite{Doke_2002}  & 19.5  & 0 & 0.83 \\  
     \hline \hline
    \end{tabular}
    \end{minipage}
    \caption{Left: values used for the band gap energies ($E_g$) and the average energy needed to produce additional electron-holes ($\varepsilon$) in the three semiconductor materials considered. Right: values used for the Thomas-Imel model for electron recoil signals in xenon and argon}
    \label{tab:semi_energies}
\end{table}

\subsection{Noble Liquid Detectors}
\label{sec:lng_yield_functions}
The Noble Liquid detectors, also known as Liquid Noble Gas detectors, used in many of the larger dark matter experiments such as XENON1T and DarkSide are primarily designed to look for nuclear recoils and treat electron recoil signals as background. The filtering of electron recoils from nuclear recoils is based on the ratio of primary (S1) to secondary (S2) photons produced via scintillation ({\it i.e.} nuclear recoils) and ionization ({\it i.e.} electron recoils) respectively. Since the S2/S1 ratio is less effective below $\sim 5$ keV most analyses and experiments stop there. However, more recently these experiments have embarked on sub-GeV DM searches utilizing the S2-only data sets~\cite{XENON:2019gfn, DarkSide:2018ppu, PandaX-II:2021nsg}. Here we argue that keeping these low energy events also allows for the detection of NSIs in a low noise noble liquid environment. 

As with semiconductors, we must convert the electron recoil energy into the detectable signal; in this case, the photoelectrons which make up the S2 signal. We adopt the Thomas-Imel model~\cite{thomas_recombination_1987} and sum the contributions from recoils of each orbital, of which the next-to-outer shells dominate. In this model, a recoiling electron with energy $E_{er}$ produces both ions $(N_i)$ and excited atoms $(N_{ex})$ with an average energy $W$, thus $E_{er} = (N_i + N_{ex})W$ since, at low energies, the fraction of ions which recombine is small $(f_R \sim 0)$ and the fraction of observable quanta is $f_e = (1-f_R)/(1+N_{ex}/N_i)\approx 0.83$. The total number of observed electrons then follows a binomial distribution of $n_1 + n_2$ with probability $f_e$ where $n_1 = \lfloor E_{er}/W\rfloor$ is the number of quanta produced immediately and $n_2$ is the number of electrons produced when photons emitted from the excited states photoionize. For a more detailed discussion of this process, see \cite{essig_first_2012}. We report the values used for $W, f_R $, and $f_e$ in Table~\ref{tab:semi_energies}.

\begin{figure}[h]
    \centering
    \includegraphics[width=0.85\textwidth]{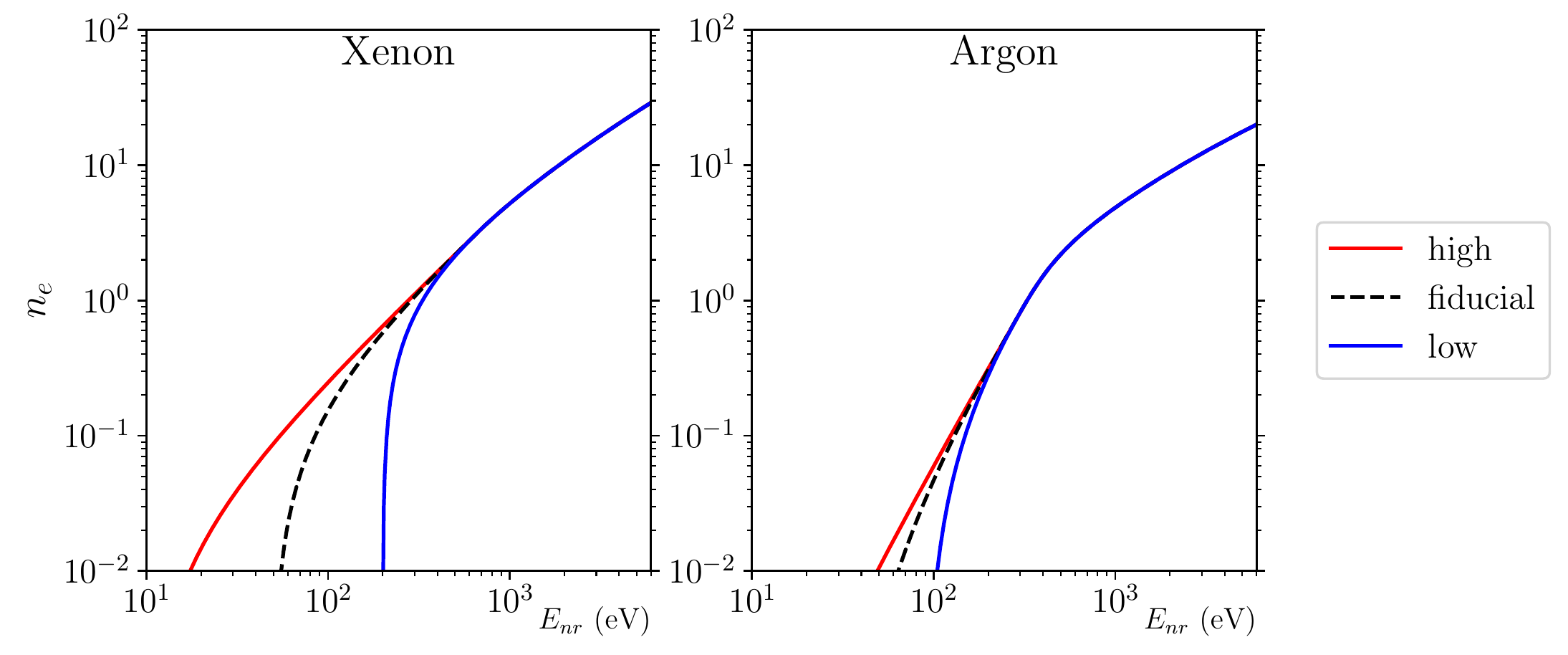}
    \caption{Charge yield of nuclear recoils in noble gases. The extrapolations to 10, 50, and 200 eV for xenon and 5, 30, and 90 eV for argon are fit to NEST simulations using a power-law}
    \label{fig:lng_y}
\end{figure}

We follow an analogous procedure to the semiconductor case for converting nuclear recoil energy into quantized charges. For liquid xenon and argon detectors, the NEST collaboration \cite{szydagis_review_2021, szydagis_nest_2011, szydagis_enhancement_2013} has provided excellent simulated models for $n_e(E_R)$ down to energies $\sim 200$ eV for nuclear recoils which we use to predict the charge yield from nuclear interactions. Below this $200$ eV limit, we again extrapolate as a power-law to a spectrum of cut-off energies for nuclear recoils and plot the results in Fig.~\ref{fig:lng_y}. As can be seen in Fig.~\ref{fig:vec_rate}, the effect of this extrapolation in nuclear recoils is limited to the $1e^-$ and $2e^-$ bins. As with silicon and germanium, \cite{sarkis_study_2020} provides an updated model for nuclear recoils in xenon, though we find that the choice of yield function has a much less significant effect in noble liquid detectors, and we use the NEST functions in this work.

Finally, to determine the event rate in each charge bin, we must consider the statistics which govern the binning of events so we use Poisson statistics to weight the recoil spectrum by its likelihood to be in each charge bin and integrate over all recoil energies to get $n_i$, the rate of observed events in the $i$-th electron bin.

\begin{equation}
    n_i = \int_{E_R^{min}}^{E_R^{max}} \frac{dR}{dE_R} \frac{n_e^i(E_R)}{i!} e^{-n_e(E_R)} dE_R
\end{equation}

\subsection{Summary and Discussion on Ionization Signals}\label{sec:yield_dep}
We provide a summary of the semiconductor and noble liquid parameters (Table~\ref{tab:semi_energies}) and a summary of the yield functions (Table~\ref{tab:yieldfnc}) used in this work.
\setlength{\tabcolsep}{0.5em} 
\renewcommand{\arraystretch}{1.3}
\begin{table}[h]
\begin{center}
\begin{tabular}{l c c c c}
\multicolumn{5}{c}{Semiconductors: $Y(E_R)$} \\
\hline
& & $E_c$ [eV] & Transition ($E_c<E_R<E_b$) & High energy model\\ \hline
\multirow{3}{*}{\makecell{\large{silicon} \\ ($E_b=600$ eV)}} &
{~~~~~~high}   & 20  & $4.49\times10^{-3}\times(E_R-E_c)^{0.486}$ & \\
& {~~fiducial} & 60  & Sarkis et al.               & Sarkis et al. \\
& {~~~~~~~low} & 200 & $1.33\times10^{-2}\times(E_R-E_c)^{0.335}$ & \\ \hline
\multirow{3}{*}{\makecell{\large{germanium} \\ ($E_b=100$ eV)}} &
{~~~~~~high}   & 5  & $7.06\times10^{-2}\times(E_R-E_c)^{0.195}$ & \\
& {~~fiducial} & 10 & Sarkis et al.             & Sarkis et al. \\
& {~~~~~~~low} & 20 & $8.35\times10^{-2}\times(E_R-E_c)^{0.164}$ & \\ \hline
\multirow{3}{*}{\makecell{\large{gallium} \\ ($E_b=254$ eV)}} &
{~~~~~~high}   & 15 & $0.183 \times(1 - e^{(E_R-E_c)/71.9)}$ & \\
& {~~fiducial} & 40 & $0.182 \times(1 - e^{(E_R-E_c)/61.5})$ & Lindhard \\
& {~~~~~~~low} & 90 & $0.180 \times(1 - e^{(E_R-E_c)/42.7})$ & \\ \hline
\multirow{3}{*}{\makecell{\large{arsenic} \\ ($E_b=254$ eV)}} &
{~~~~~~high}   & 15 & $0.180 \times(1 - e^{(E_R-E_c)/71.9})$ & \\
& {~~fiducial} & 40 & $0.179 \times(1 - e^{(E_R-E_c)/61.5})$ & Lindhard \\
& {~~~~~~~low} & 90 & $0.177 \times(1 - e^{(E_R-E_c)/42.7})$ & \\ \hline
\multicolumn{5}{c}{\makecell{ \\ Noble Liquids: $n_e(E_R)$}} \\ \hline
\multirow{3}{*}{\makecell{\large{xenon} \\ ($E_b=667$ eV)}} &
{~~~~~~high}   & 10  & $7.61\times10^{-4}\times(E_R-E_c)^{1.29}$ & \\
& {~~fiducial} & 50  & $1.36\times10^{-3}\times(E_R-E_c)^{1.21}$ & NEST \\
& {~~~~~~~low} & 200 & $1.16\times10^{-2}\times(E_R-E_c)^{0.914}$& \\ \hline
\multirow{3}{*}{\makecell{\large{argon} \\ ($E_b=300$ eV)}} &
{~~~~~~high}   & 10  & $1.57\times10^{-6}\times(E_R-E_c)^{2.31}$ & \\
& {~~fiducial} & 50  & $5.70\times10^{-6}\times(E_R-E_c)^{2.12}$ & NEST \\
& {~~~~~~~low} & 200 & $1.16\times10^{-4}\times(E_R-E_c)^{1.66}$ & \\ \hline
\end{tabular}
\end{center}
\caption{Yield functions for semiconductors ($E_e = E_R\times Y(E_R)$) and noble liquids ($n_e(E_R)$). Yields are zero below the cut-off ($E_c$) and match the high-energy model above the branching energy $E_b$. The high-energy models are given by Sarkis et al.~\cite{sarkis_study_2020,Sarkis:2021agv}, Lindhard~\cite{lindhard1963integral}, and NEST~\cite{szydagis_nest_2011}}
\label{tab:yieldfnc}
\end{table}%
The dependence of Eq.~\eqref{diff_rate} on the derivative of the yield function as well at the shift of integration limits means that a more gradually changing yield function will significantly enhance the observed rate in the detector. We highlight the effect of the latter dependency on the $1e^-$ bin in GaAs with the shaded bars in Fig.~\ref{fig:yield}. The width of these bars correspond to the range of nuclear recoil energies which contribute to each charge bin and their effect can be seen in our results that follow. Here, we see that the rates in the lowest charge bins can vary by orders of magnitude between yield functions highlighting the need for accurate, empirical measurements of the yield function at low energy. We investigate the effects of the choice of yield functions in more detail in Appendix~\ref{app:yield}

\section{Results} \label{sec:results}
\begin{figure}[t]
    \centering
    \includegraphics[width=\textwidth]{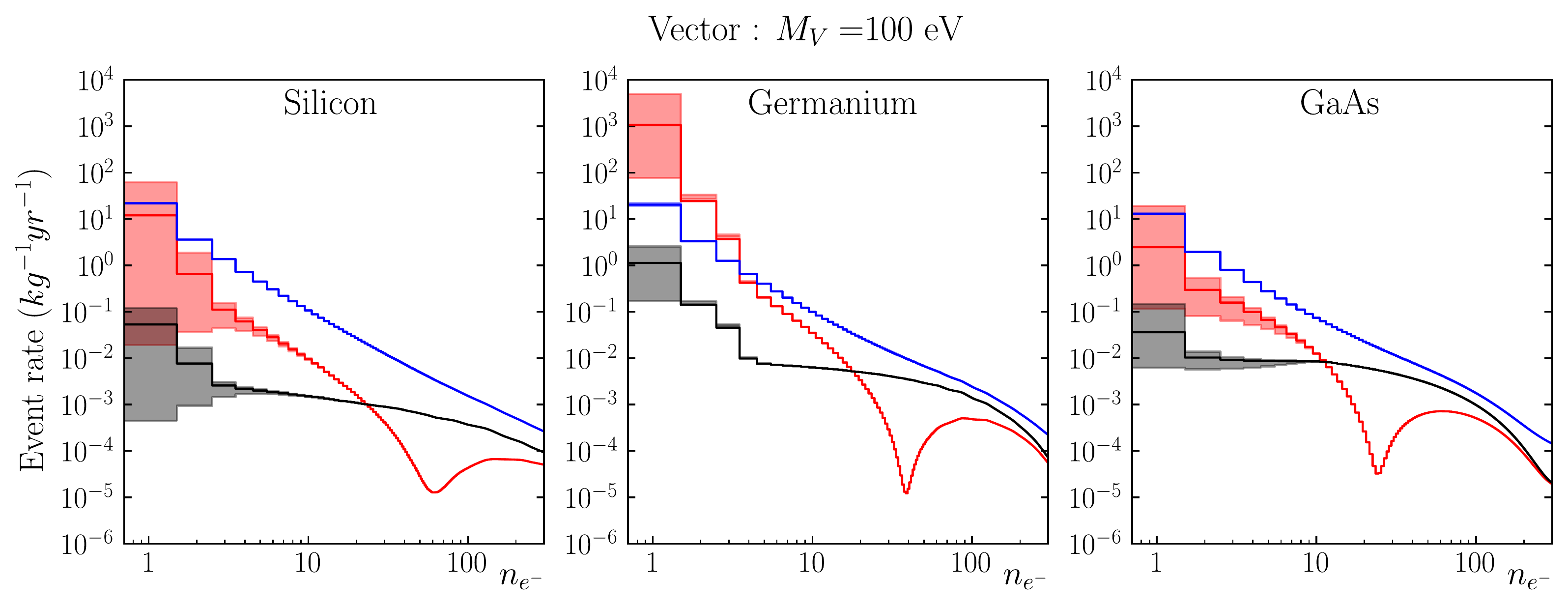}
    \raggedright
    \includegraphics[width=0.94\textwidth]{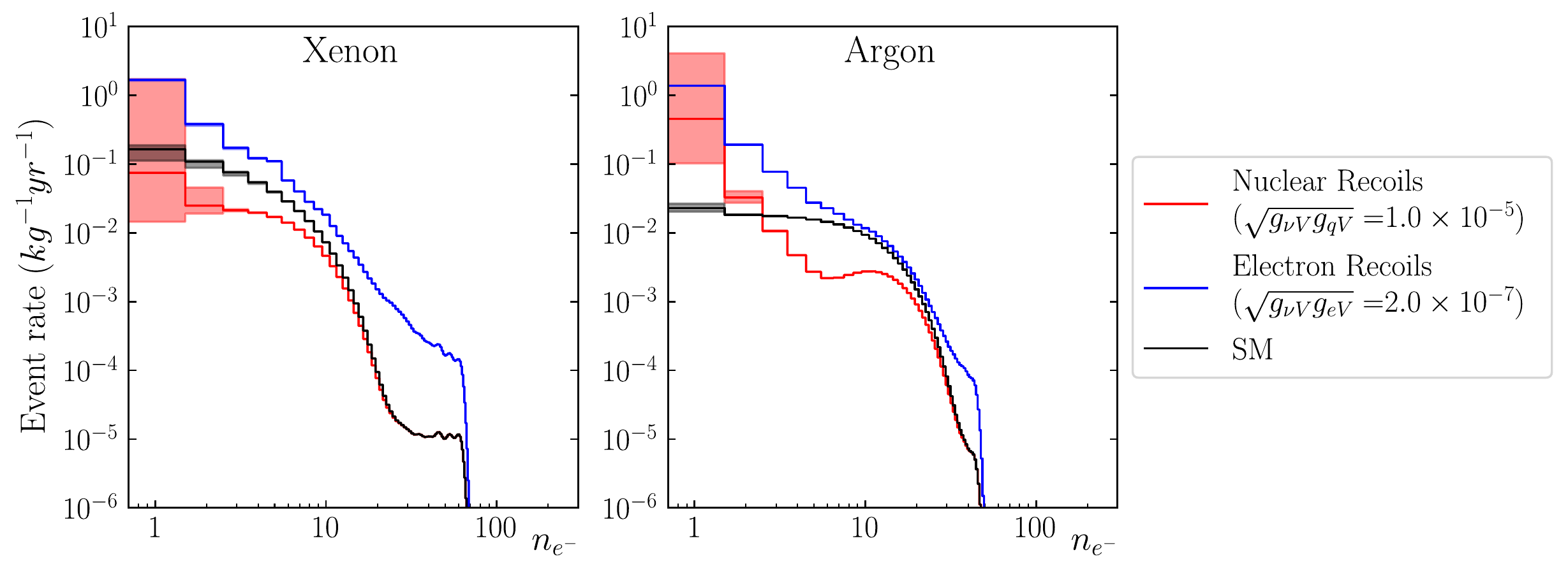}
    \caption{Event rates from nuclear (red) and electron (blue) recoil from a 100 eV vector mediator near the projected bound for vIOLETA \cite{fernandez-moroni_physics_2021}. The SM rate is given in black. The bands around the nuclear and SM rates come from varying the choice of ionization yield function.}
    \label{fig:vec_rate}
\end{figure}

In this section, we present the results for the scattering rates as well as projections on the constraints of the NSI couplings. The simplified models -- vector, scalar, axial vector, pseudoscalar -- contain three free parameters: the mediator mass, the nuclear coupling, and the electron coupling. For the neutrino magnetic moment model, there is only one free parameter: the neutrino magnetic moment $\mu_\nu$. 
In what follows, we present our results in terms of the free parameters of the models under consideration.
Note that for the simplified models, the nucleon and electron couplings are independent and must be constrained separately while the one parameter model for the magnetic moment allows for a more detailed analysis which we will discuss in Sec.~\ref{sec:magmoment}. To generate constraints on each model from semiconductors and noble liquid detectors, we consider a detector which sees no events above the predicted background at 90\%C.L., {\it e.g.} 3 events. 

For the simplified models, we present the 90\%C.L. projected constraints at exposures of 0.1, 1, and 30 kg-yrs in semiconductors which corresponds to the expected exposures of SENSEI, DAMIC-M, and OSCURA~\cite{baxter_status_2020}. We consider exposures of 20, 200, and 1000 tonne-yrs in xenon, corresponding to XENONnT, Darwin \cite{darwin_collaboration_darwin_2019}, and a more far future detector and 50, 360, and 1000 tonne-yrs in argon, the first two being roughly the exposure of DarkSide-20k and ARGO \cite{giovanetti_2020}, and the third is again a far future detector. Because of their larger exposures, the noble liquid detectors can place a tighter bound on the electron couplings, but the lower threshold of the semiconductors makes them more sensitive to CEvNS and hence improves the quark coupling sensitivity.

As mentioned in Sec.~\ref{sec:scattering}, we neglected the effects of binding energy in the electron scattering cross-sections. They restrict the available phase space for the interaction and reduce the cross-section at low energies by a factor $\sim 3$. We approximated this effect in xenon with a series of step functions as in \cite{CHEN2017656} which led to a reduction in sensitivity of $\mathcal{O}(30\%)$ for vector models and $\mathcal{O}(20\%)$ in scalar models, both in the low mass limit. We expect the magnetic moment sensitivity to behave like the scalar under this correction due to its similar $\frac{1}{E_R}$ dependence at low energies and also expect a similar effect in other materials In what follows, we present the results without the effects of binding energy, leaving a more rigorous analysis for later work.

\subsection{Simplified Models}\label{sec:simplified}

\begin{figure}[t]
    \includegraphics[width=\textwidth]{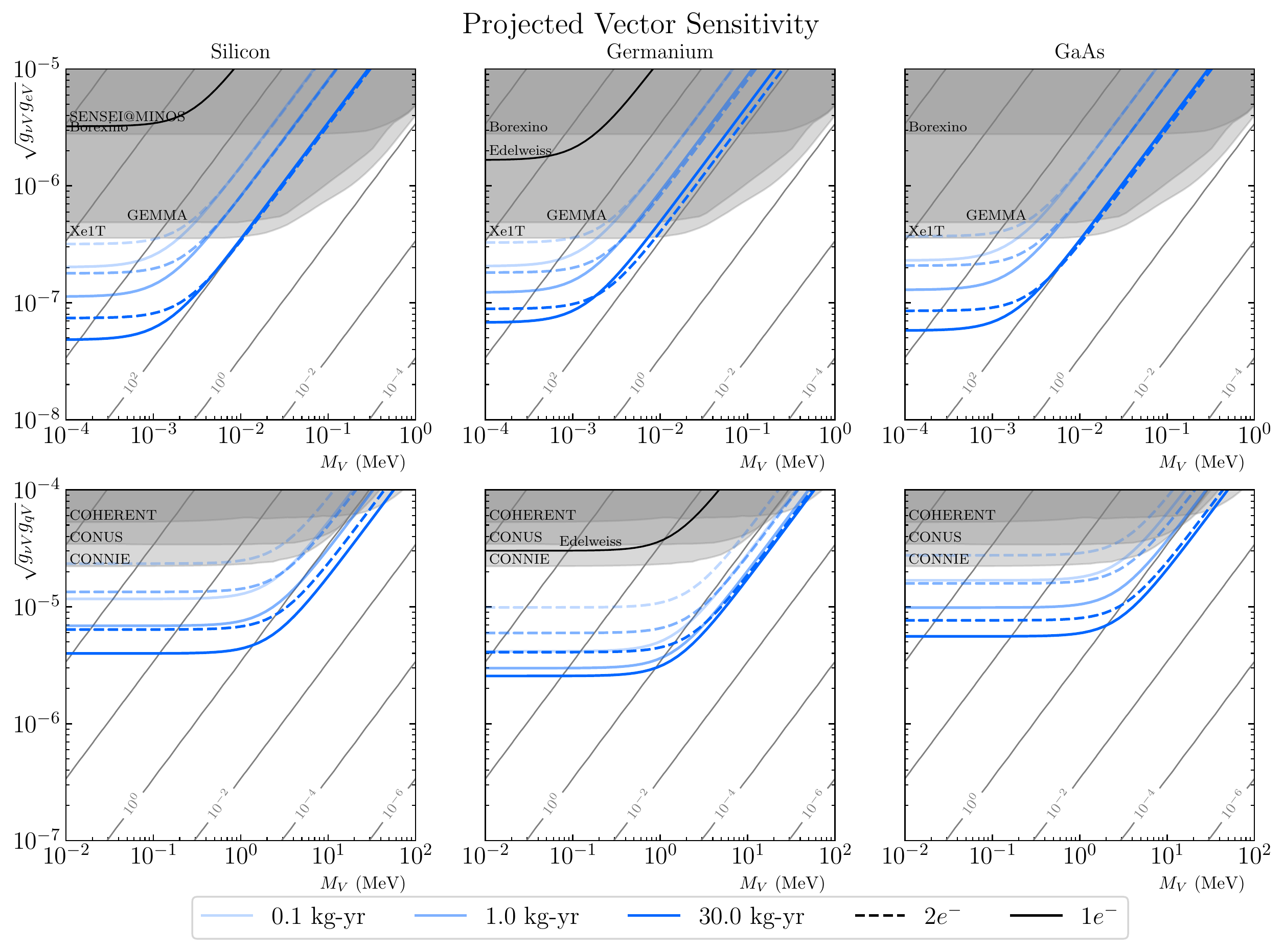}
    \caption{Sensitivity to NSIs mediated by a new vector in Si (left), Ge (middle), and GaAs (right) for electron (top) and nuclear (bottom) couplings. The line shading indicate the exposures of 0.1 (lightest), 1, and 30 (darkest) kg-years, while the solid (dashed) lines denote the one (two) electron bin. Also included are the current constraints from SENSEI@MINOS~\cite{sensei_collaboration_sensei_2020} in silicon and Edelweiss~\cite{edelweiss_collaboration_first_2020} in germanium. The shaded gray regions are the existing constraints on the parameter space and the grey contours are constant $\epsilon$ as defined in Eq. \eqref{eq:coupling_rel}. See text for details.}
    \label{fig:vector_semi_sens}
\end{figure}

\begin{figure}[t]
    \centering
    \includegraphics[width=0.85\textwidth]{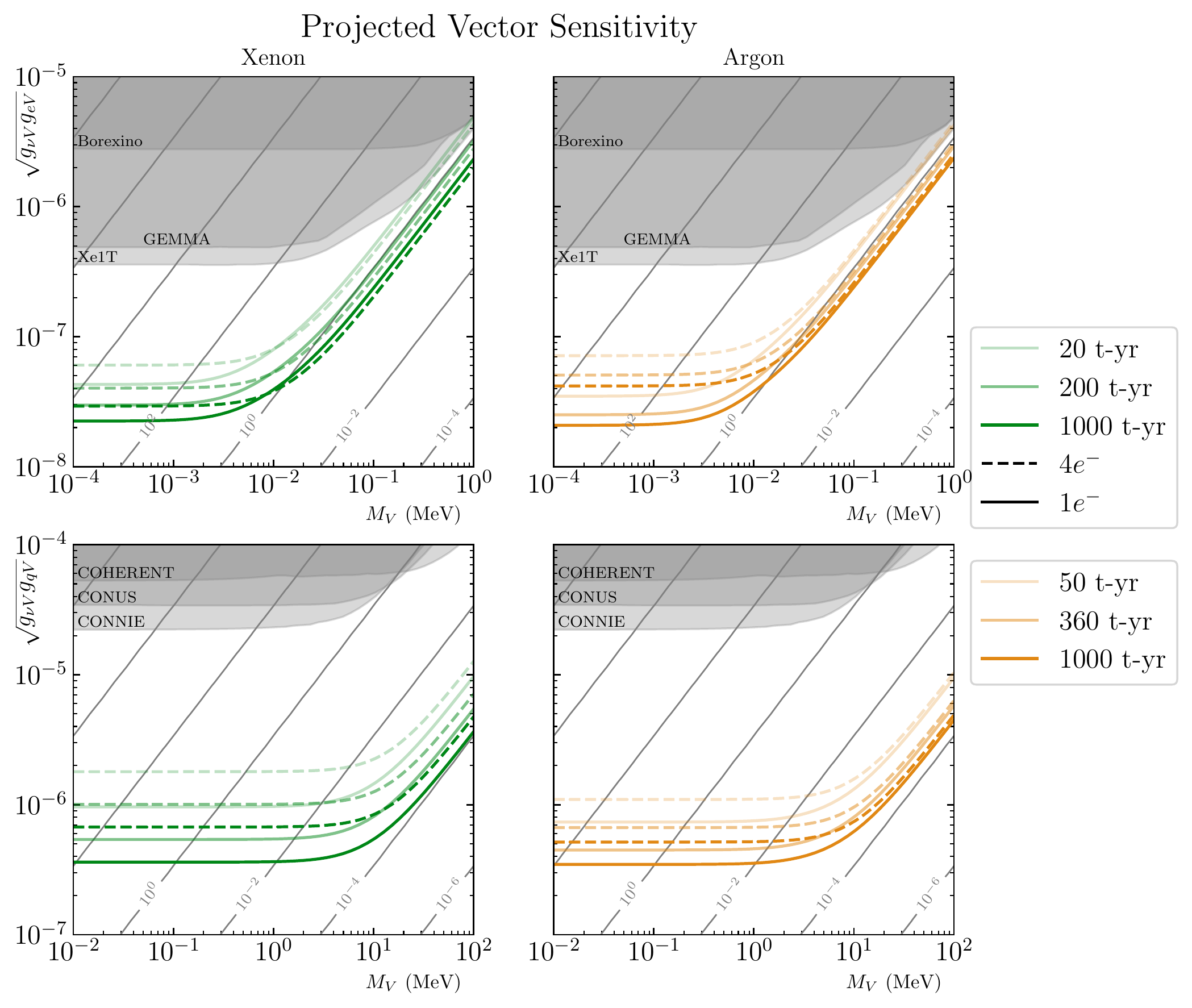}
    \caption{Sensitivity to NSIs mediated by a new vector in Xe (left) and Ar (right) for electron (top) and nuclear (bottom) couplings. The various shadings for the green (orange) lines denote the different exposures for xenon (argon), while the solid (dashed) lines denote the one (four) electron threshold. The shaded gray regions are the existing constraints on the parameter space and the grey contours represent constant $\epsilon$ as defined in Eq. \eqref{eq:coupling_rel}. See text for details.}
    \label{fig:vector_lng_sens}
\end{figure}

Using the yield functions from Table~\ref{tab:yieldfnc}, the methods described in Sec.~\ref{sec:scattering}, and the fluxes from Fig.~\ref{fig:solarnuflux} we are able to calculate the event rate for an interaction mediated by either a scalar, pseudoscalar, vector, or axial vector coupling either to nucleons or electrons. In the following subsections, we plot illustrative examples of these rates for 100 eV mediators with generation independent couplings near the sensitivity projected for the vIOLETA experiment~\cite{fernandez-moroni_physics_2021} for both electron (red) and nuclear (blue) recoils. Note that the NSI rates are the \emph{combined} SM background (CEvNS and electron recoils) with the NSI contribution. As such, the region where the blue line is below the black in the vector model indicates destructive interference and a negative NSI contribution. The shading indicates the uncertainties resulting from the choice of yield function. Although these only effect the nuclear recoils, they are present in the background from SM CEvNS so we keep them for calculations of our sensitivity to electron couplings though they are too small to be visible in Fig.~\ref{fig:vec_rate}. Both the electron and nuclear rates are sensitive to uncertainties in the solar flux, but these uncertainties are subdominant to the choice of yields function. As such, the majority of the uncertainties in the event rate can be improved without changing the detectors, but rather improved measurement or modeling of the yield function.

\begin{figure}[h]
    \centering
    \includegraphics[width=\textwidth]{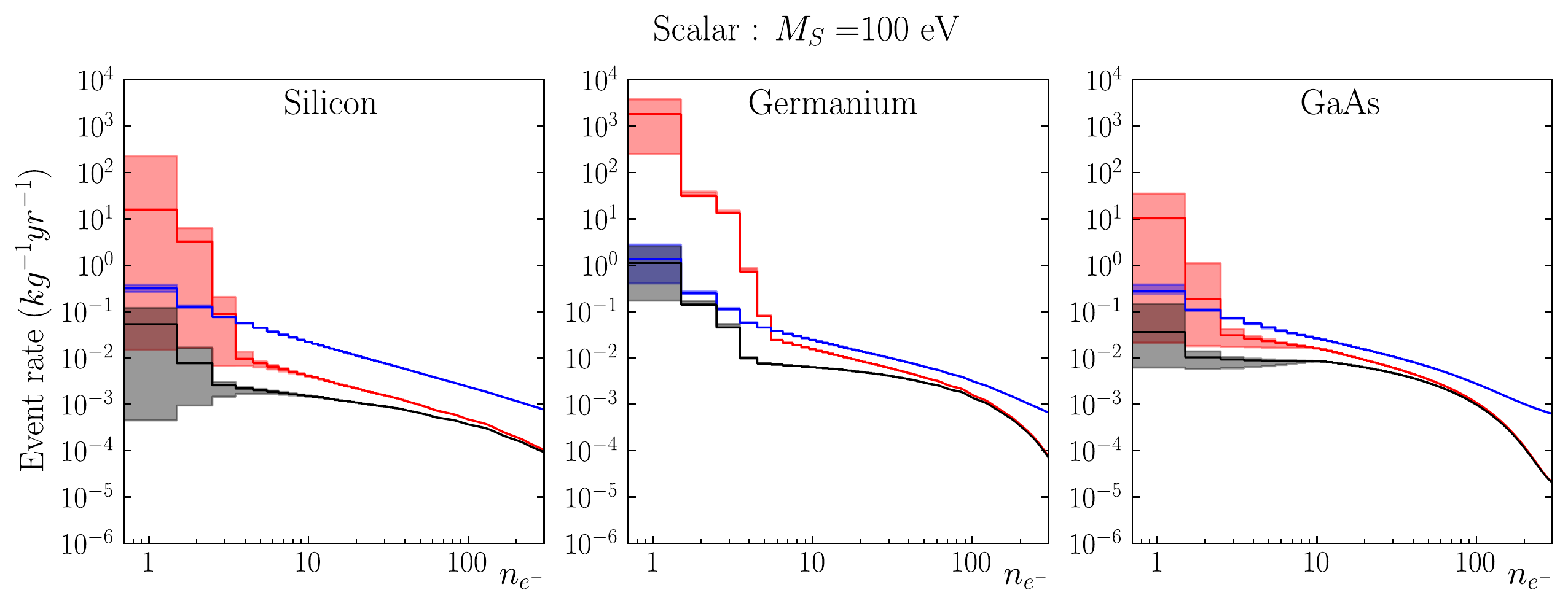}
    \raggedright
    \includegraphics[width=0.94\textwidth]{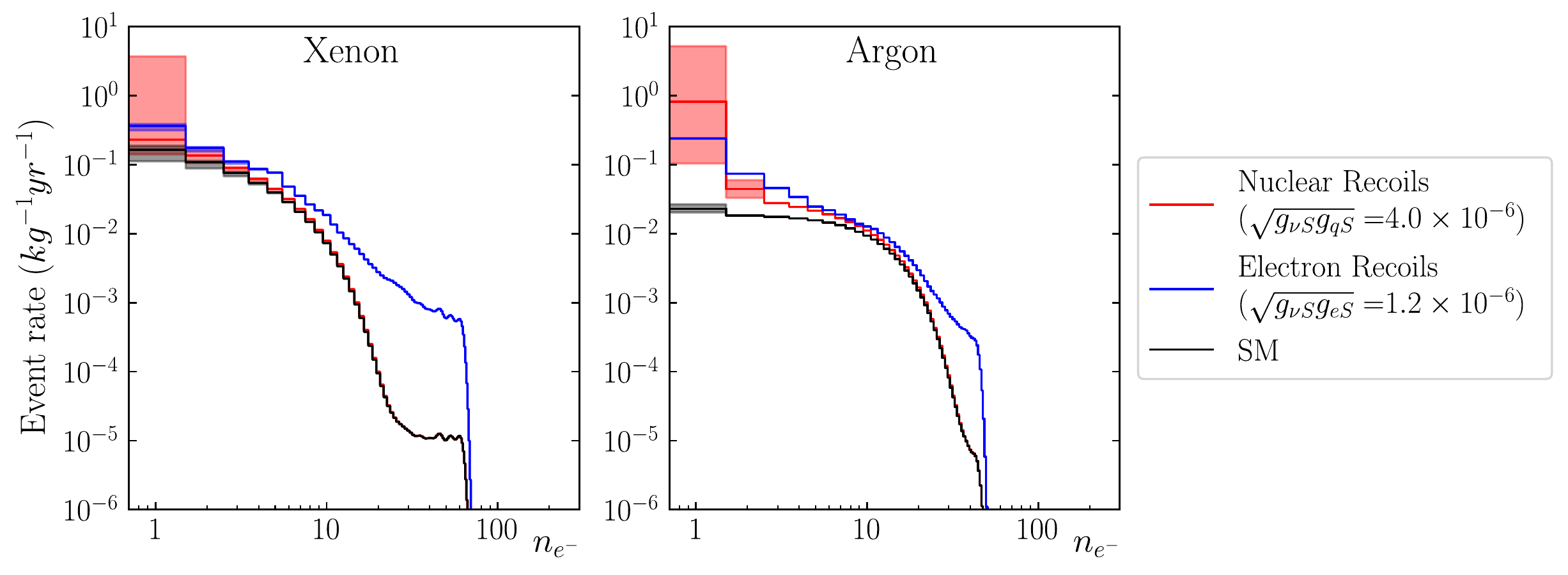}
    \caption{Event rates from nuclear (red) and electron (blue) recoil from a 100 eV scalar mediator near the projected bound for vIOLETA \cite{fernandez-moroni_physics_2021}. The SM rate is given in black. The bands around the nuclear and SM rates come from varying the choice of ionization yield function.}
    \label{fig:scal_rate}
\end{figure}

\begin{figure}[h]
    \centering
    \includegraphics[width=\textwidth]{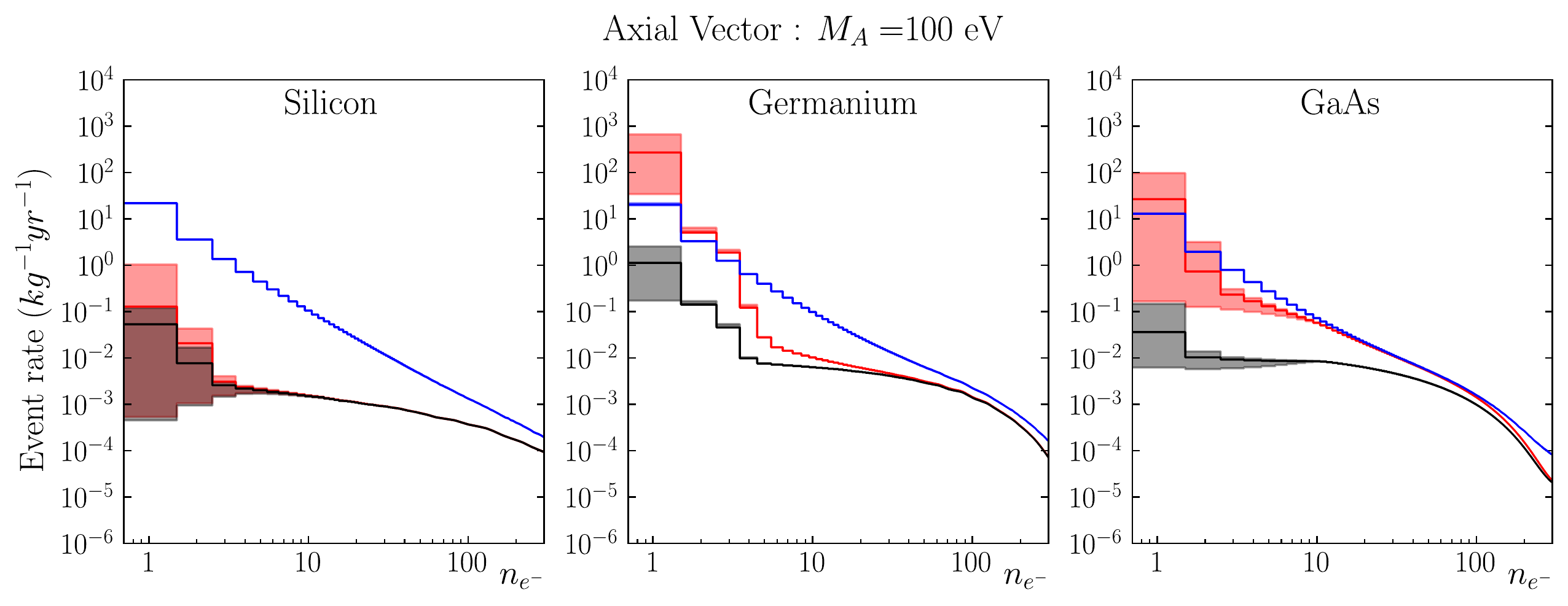}
    \raggedright
    \includegraphics[width=0.94\textwidth]{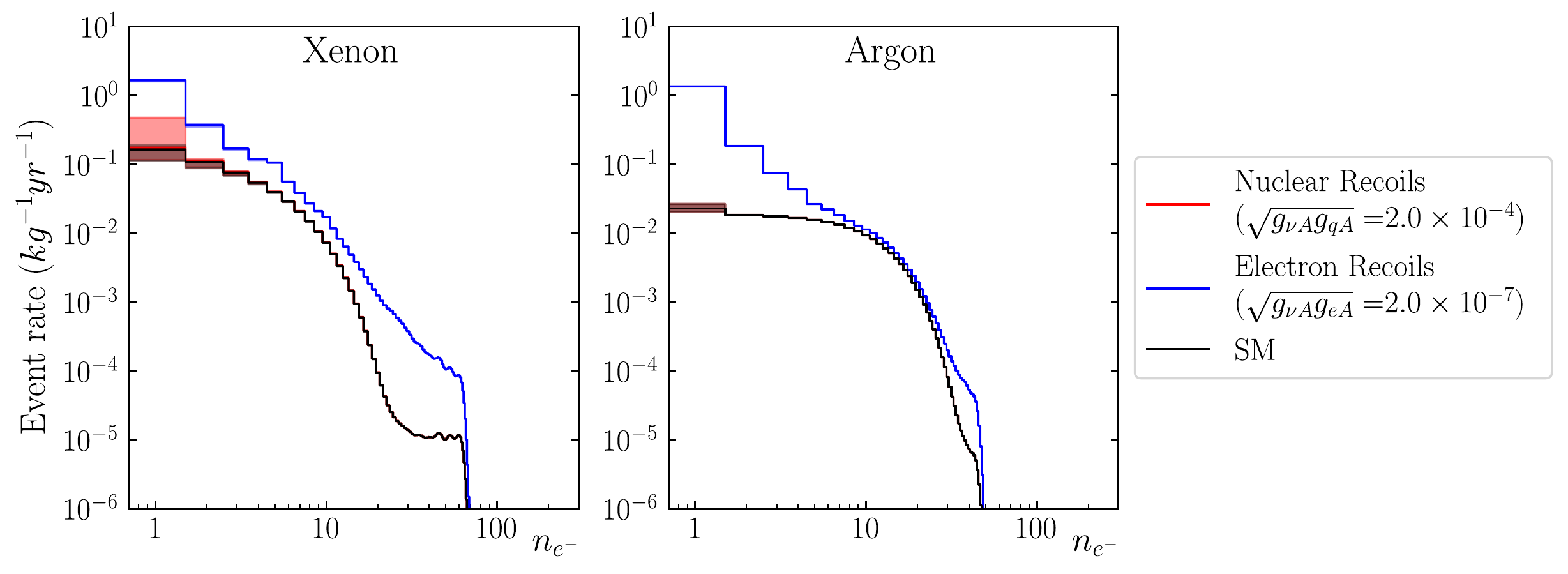}
    \caption{Event rates from nuclear (red) and electron (blue) recoil from a 100 eV axial vector mediator near the projected bound for vIOLETA \cite{fernandez-moroni_physics_2021}. The SM rate is given in black. The bands around the nuclear and SM rates come from varying the choice of ionization yield function.}
    \label{fig:axial_rate}
\end{figure}

Each of the bins in Figs.~\ref{fig:vec_rate}, \ref{fig:scal_rate}, \ref{fig:axial_rate}, and \ref{fig:pseudo_rate} can be used to place a constraint on the parameters of the model used to generate the NSI rate. At the exposures of interest in semiconductors, the background is essentially undetectable and any NSI signal will be dominated by the 1 and 2 $e^-$ bins since they correspond to the lowest energy recoils, so we focus on those. The $1e^-$ bin will generally provide the stronger constraint at low exposures, but the $1e^-$ bin may still have too much noise in some experimental environments.

We consider two simplified cases in noble liquid detectors. The first is a sum of all events with four or greater electrons as such detectors are already sensitive to. The second is a sum of all events including those with less than four electrons. In each of these cases, the sensitivity could be improved by evaluating a likelihood function with a global fit to the entire set of event rates, but we take the simplified cases as conservative limits and simply note that qualitatively, the more robust analysis would have a larger improvement for the all-charges-included case since the low charge behavior has a stronger deviation from the SM and switches between constructive and destructive interference in the vector mediated case. In the vector case, we note that we do not consider the direction of the deviation from the SM, but rather consider the absolute value of this deviation and its ratio with the SM expectation.

\begin{figure}[t]
    \centering
    \includegraphics[width=\textwidth]{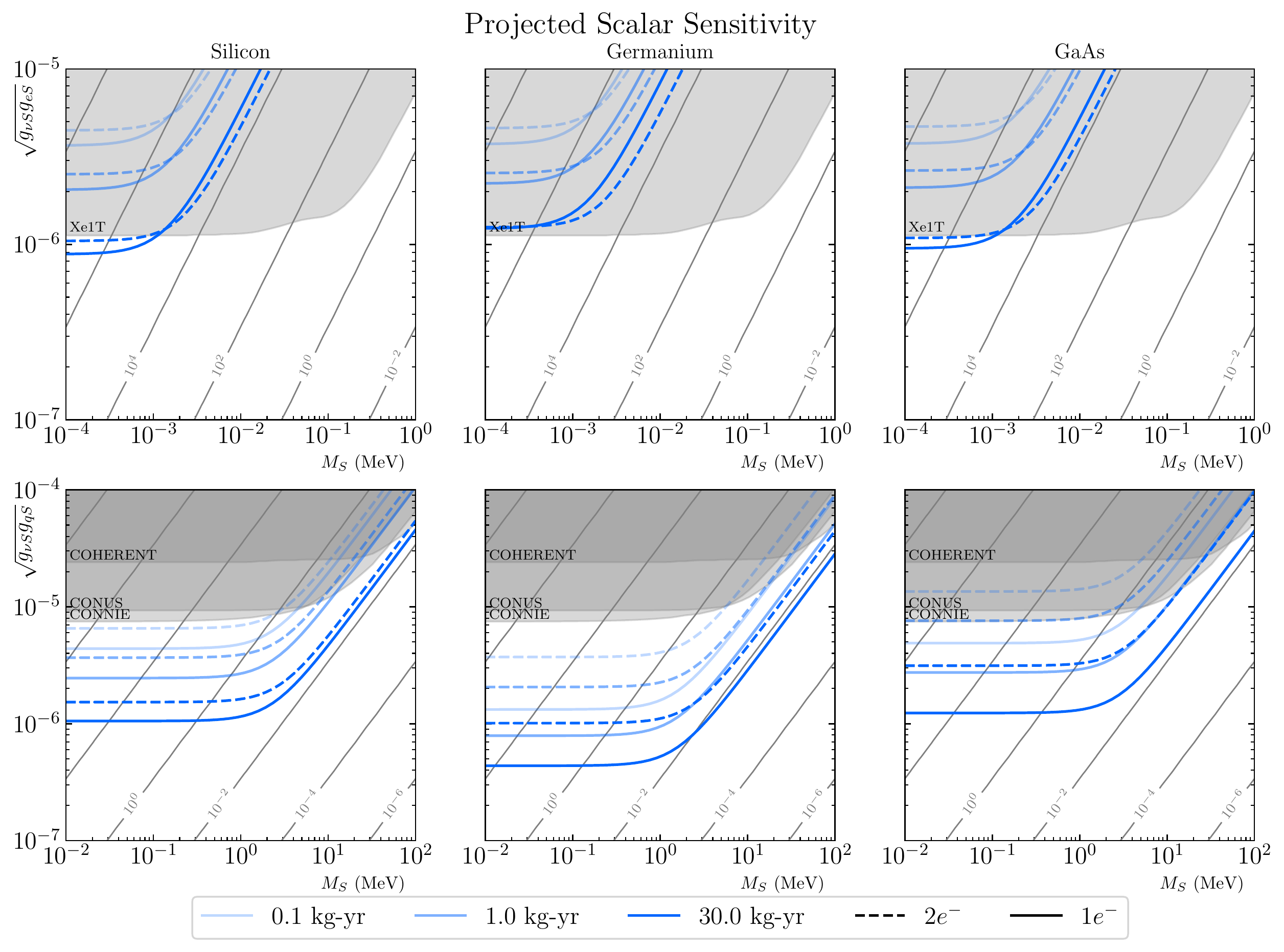}
    \caption{Sensitivity to NSIs mediated by a new scalar in Si (left), Ge (middle), and GaAs (right) for electron (top) and nuclear (bottom) couplings. The line shading indicates the exposures of 0.1 (lightest), 1, and 30 (darkest) kg-years, while the solid (dashed) lines denote the one (two) electron bin. The shaded gray regions are the existing constraints on the parameter space. The grey contours represent constant $\epsilon$ as defined in Eq. \eqref{eq:coupling_rel}. See text for details.}
    \label{fig:scalar_semi_sens}
\end{figure}

In the subsections that follow, we overlay our projected sensitivity for the materials discussed in Sec.~\ref{sec:detectors} on existing constraints. Existing nuclear coupling constraints come from CONUS~\cite{conuscollaboration2021novel}, CONNIE~\cite{aguilar-arevalo_search_2020}, and COHERENT~\cite{Liao_coherent, B_hm_2019, Gonzalez-Garcia:2018dep, AristizabalSierra:2019ykk, AristizabalSierra:2021kht}. The latter is a collection of detectors using liquid argon, CsI(Na), and NaI(Tl) at the Spallation Neutron Source at Oakridge National Laboratory~\cite{akimov_observation_2017} while the first two are germanium and silicon detectors at commercial nuclear power plants. As such, they are only sensitive to couplings to electron-neutrinos while COHERENT is sensitive to all flavors. Their constraints also depend on the particular properties of their detector materials. For instance, the argon detector subsystem of COHERENT cannot constrain axial vector mediators since argon nuclei have no spin. While CONUS and CONNIE can place constraints on the coupling to electrons, they are subdominant to some larger experiments. We focus instead on the results of XENON1T~\cite{boehm_light_2020, AristizabalSierra:2020edu, AristizabalSierra:2017joc}, Borexino~\cite{harnik_exploring_2012}, Texono~\cite{texono}, and GEMMA~\cite{fernandez-moroni_physics_2021}. As a dark matter detector, XENON1T is designed to search for DM-nucleus recoils which are individually indistinguishable from CEvNS. Such a distinction could only be made by the energy distribution of those recoils. The most distinct feature of that distribution is the much lower energy of CEvNS. Since DM recoils are at higher energy, XENON1T is not optimized to see CEvNS so we can only place competitive constraints on electron couplings since the lighter target recoils with higher energy. Borexino~\cite{Borexino:2008dzn} uses a 5-ton liquid scintillator to detect scattering of solar neutrinos off of electrons. It is therefore sensitive to couplings with any neutrino flavor. Texono~\cite{texono} and GEMMA~\cite{beda_gemma_2013} are reactor experiments, with the former using a 187 kg CsI(Ti) scintillating crystal and the latter using a 1.5 kg germanium detector. As with other reactor experiments, they are only sensitive to electron-neutrinos.

\begin{equation}\label{eq:coupling_rel}
    \epsilon = \frac{g_{\nu i}g_{SM i}}{M_i^2G_F}
\end{equation}

As discussed in Sec.~\ref{sec:intro} we are using a different convention than much of the literature regarding NSIs. Elsewhere, the NSI strength is often given by $\epsilon$ (see {\it e.g.} \cite{dev_neutrino_2019}) which parameterizes the strength relative to the SM weak force and relates to our couplings by Eq. \eqref{eq:coupling_rel} where $g_{\nu i}$ and $g_{SM i}$ are the couplings from Table~\ref{tab:cross_sections} coupling the mediator ($i$) to neutrinos and quarks/electrons respectively. To visualize this relation, we include contours of constant $\epsilon$ in the figures covering the relevant parameter space.
\subsubsection{Vector}\label{sec:vector}
\begin{figure}
    \centering
    \includegraphics[width=0.85\textwidth]{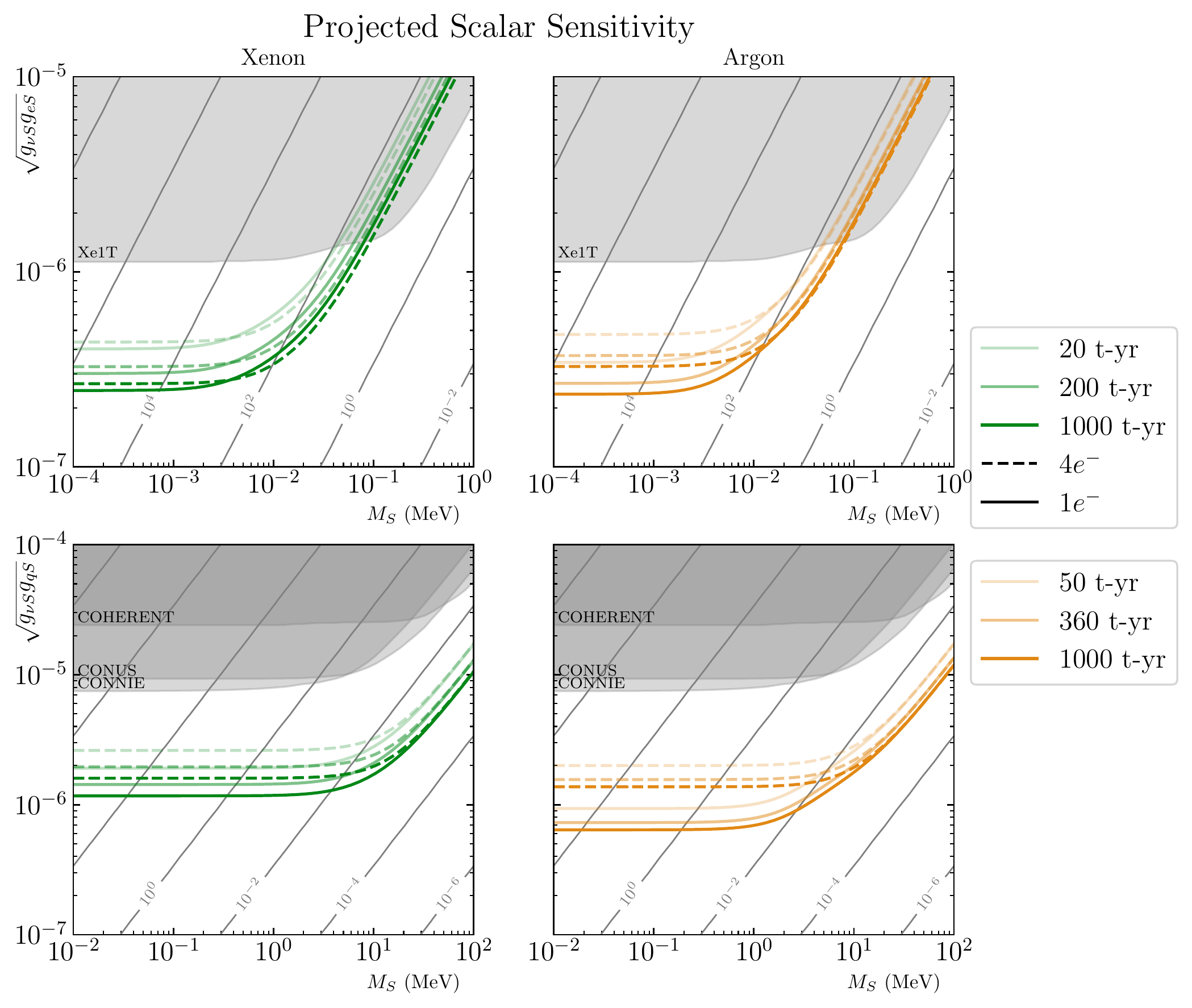}
    \caption{Sensitivity to NSIs mediated by a new scalar in Xe (left) and Ar (right) for electron (top) and nuclear (bottom) couplings. The various shadings for the green (orange) lines denote the different exposures for xenon (argon), while the solid (dashed) lines denote the one (four) electron threshold. The shaded gray regions are the existing constraints on the parameter space. The grey contours represent constant $\epsilon$ as defined in Eq. \eqref{eq:coupling_rel}. See text for details.}
    \label{fig:scalar_lng_sens}
\end{figure}
We show the rates for neutrino-nucleus and neutrino-electron scattering in Fig.~\ref{fig:vec_rate} for $m_V=100$ eV and couplings near the projected bound for the vIOLETA experiment~\cite{fernandez-moroni_physics_2021}. Here, we clearly see the enhancement in the rate at low-energies that result from the $\sim1/E_R^2$ scaling for $E_R>m_V^2/m_T$, as discussed in Sec.~\ref{NSI_rates}. In addition, the new vector mediator leads to a unique feature in the recoil spectrum, which arises from the destructive interference between the new vector and the SM weak bosons. This destructive interference is visible at $n_e\sim 20-40$ in the semiconductor materials, where the nuclear recoil event rate drops below the SM rate in Fig.~\ref{fig:vec_rate}. Although not pertinent for our studies, which focus on the lowest charge bins, this unique feature could serve as a powerful means of discriminating between signal and background.

In Figs.~\ref{fig:vector_semi_sens} and~\ref{fig:vector_lng_sens}, we plot the projected sensitivities to a vector mediator in semiconductor and noble liquid detectors, respectively. In Ar and Xe, the $1 (4) e^-$ refers to the lower bound of the summation of events while in the semiconductors, the $1 (2) e^-$ refers explicitly to the charge bin used to derive the constraints. Electron-coupling constraints are plotted in the upper panels and nuclear ones on the lower. We see that, despite the destructive interference that appears at the high-electron bin, the low-thresholds of these experiments lead to an improvement of the sensitivity to both electron and nuclear couplings. In particular, the improvement from decreasing the threshold can, in some cases, be more significant than increasing the exposure. For comparison, we also calculate the constraints from SENSEI@MINOS~\cite{sensei_collaboration_sensei_2020} -- the results from CDMS-HVeV~\cite{supercdms_collaboration_constraints_2020} will be comparable -- and Edelweiss~\cite{edelweiss_collaboration_first_2020} on the silicon and germanium panels, respectively, and show that they can be comparable to some of the existing constraints for low vector masses. 

\subsubsection{Scalar}\label{sec:scalar}
\begin{figure}[h]
    \centering
    \includegraphics[width=\textwidth]{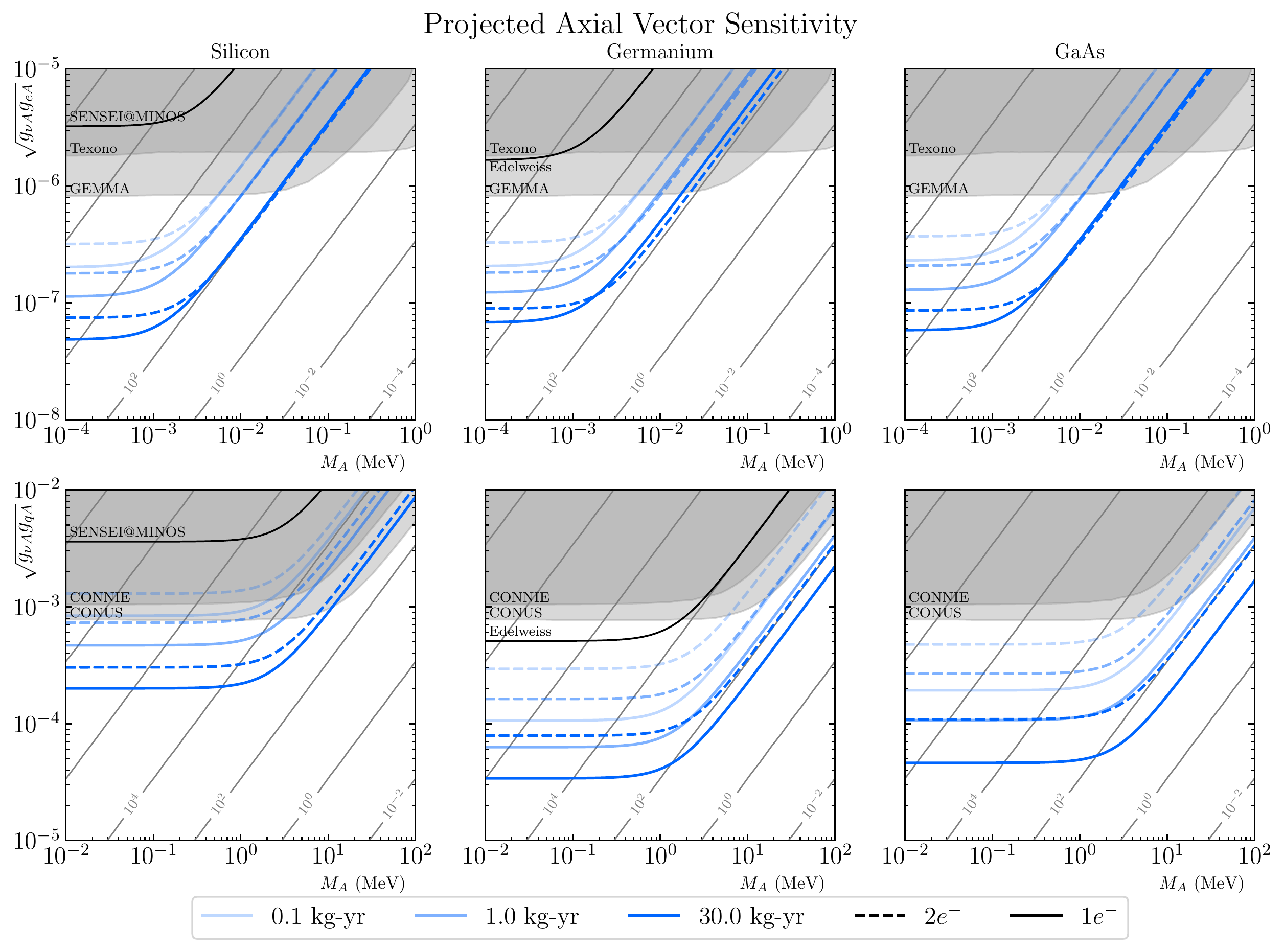}
    \caption{Sensitivity to NSIs mediated by a new axial vector in Si (left), Ge (middle), and GaAs (right) for electron (top) and nuclear (bottom) couplings. The line shading indicates the exposures of 0.1 (lightest), 1, and 30 (darkest) kg-years, while the solid (dashed) lines denote the one (two) electron bin. In the lower left panel, we see the effect from the suppression due to the small abundance-averaged spin in silicon. The shaded gray regions are the existing constraints on the parameter space. The grey contours represent constant $\epsilon$ as defined in Eq. \eqref{eq:coupling_rel}. See text for details.}
    \label{fig:axial_semi_sens}
\end{figure}

\begin{figure}[h]
    \centering
    \includegraphics[width=0.85\textwidth]{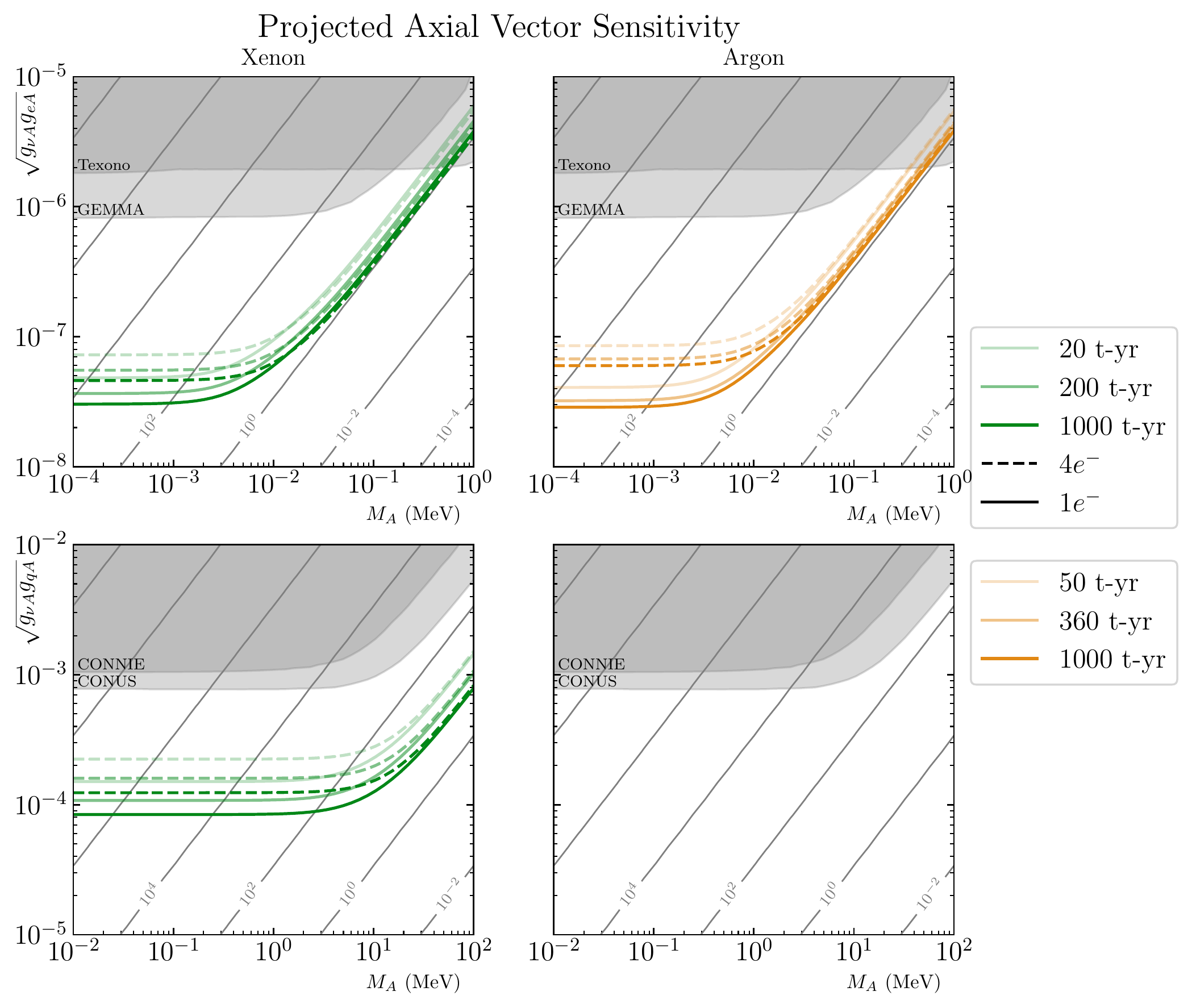}
    \caption{Sensitivity to NSIs mediated by a new axial vector in Xe (left) and Ar (right) for electron (top) and nuclear (bottom) couplings. The various shadings for the green (orange) lines denote the different exposures for xenon (argon), while the solid (dashed) lines denote the one (four) electron threshold. The lower right panel shows the effects of argon's lack of net nuclear spin. The shaded gray regions are the existing constraints on the parameter space. The grey contours represent constant $\epsilon$ as defined in Eq. \eqref{eq:coupling_rel}. See text for details.}
    \label{fig:axial_lng_sens}
\end{figure}
We plot an example event rate for the scalar mediated interaction in Fig.~\ref{fig:scal_rate}. In contrast to the vector case discussed above, we see that the scalar recoil rate has a weaker enhancement at low $E_R$, as discussed in Sec.~\ref{NSI_rates}. Thus, we expect the sensitivity of semiconductor targets to be reduced as these targets have relatively small exposures and rely on the enhancement at low-energies to have competitive bounds. This reduced sensitivity is evident in Figs.~\ref{fig:scalar_semi_sens} and ~\ref{fig:scalar_lng_sens} where even a 30 kg-yr semiconductor struggles to match the results of XENON1T. The new constraints from SENSEI@MINOS and Edelweiss are $4.0\times10^{-5}/1.3\times10^{-4}$ and $3.0\times10^{-5}/1.1\times10^{-4}$ respectively for the electron/quark couplings; these values are not competitive with existing constraints and thus do not appear in Fig.~\ref{fig:scalar_semi_sens}.

\subsubsection{Axial vector}\label{sec:axial}
Like the vector model, the axial vector mediated cross-section has a $1/E_R^2$ enhancement for light mediators at low energies so semiconductors regain their advantage. However, there is an additional suppression due to the coupling of the axial vector to spin, which we can see explicitly in the lower panels of Figs.~\ref{fig:axial_semi_sens} and~\ref{fig:axial_lng_sens}. Of particular note is argon and silicon; argon is not sensitive to axial vector mediated nuclear scattering since it has no nuclear spin (the same is true of $^4$He although we do not include it in this work). This feature of argon, combined with its sensitivity to the other models, makes it a useful material to distinguish an axial vector mediator if a signal is seen in other materials. Similarly, the abundance-averaged spin of silicon is small, so the rate is also suppressed there.
\begin{figure}[h]
    \centering
    \includegraphics[width=\textwidth]{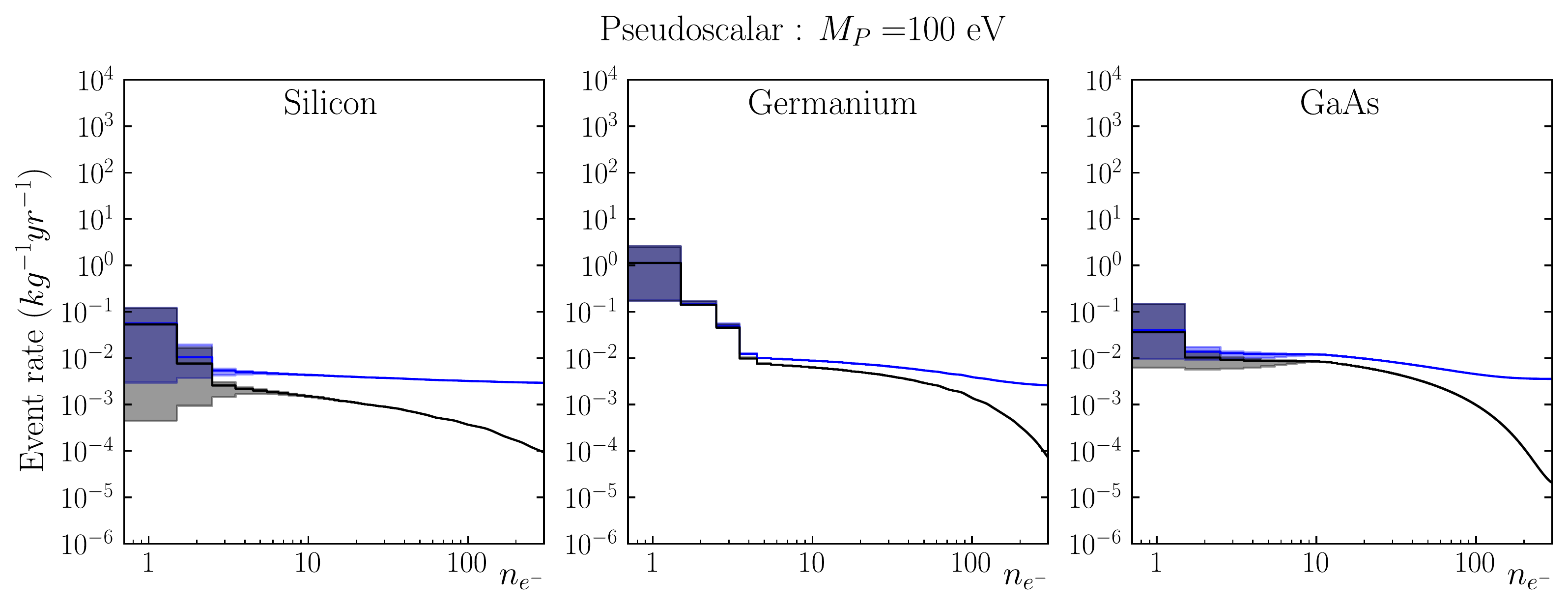}
    \raggedright
    \includegraphics[width=0.94\textwidth]{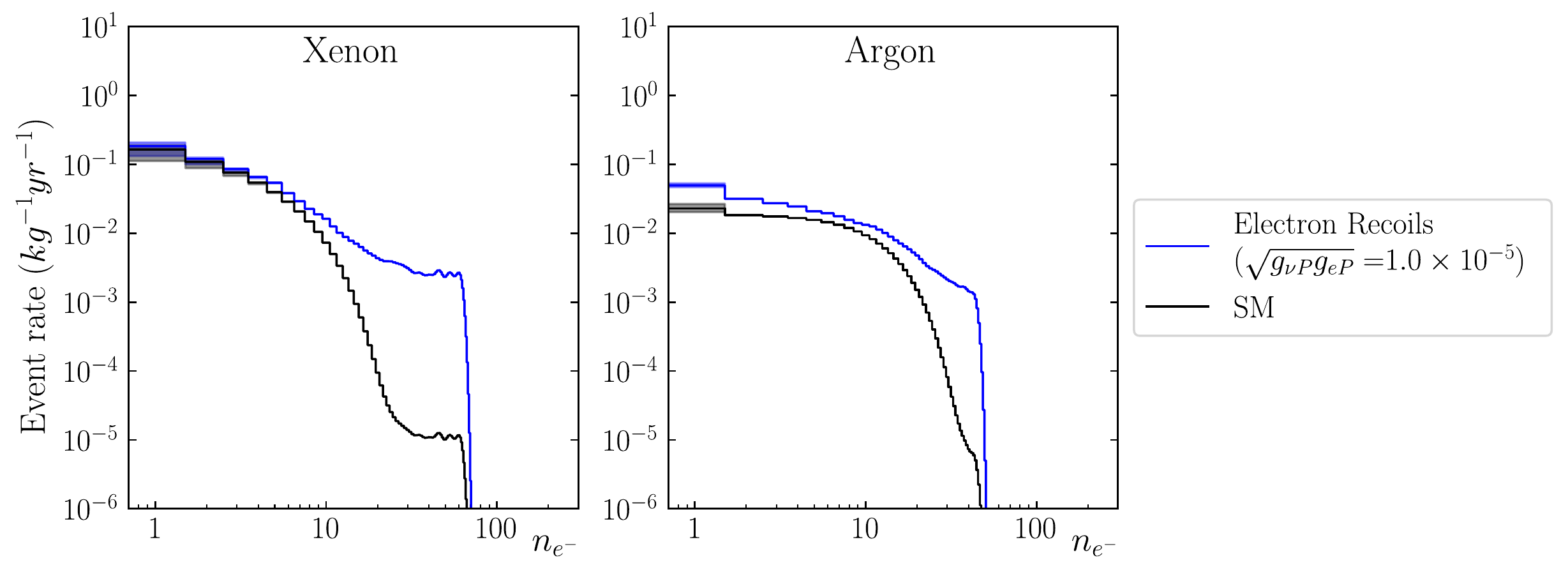}
    \caption{Event rates from electron (blue) recoil from a 100 eV pseudoscalar near the projected bound for vIOLETA \cite{fernandez-moroni_physics_2021}.}
    \label{fig:pseudo_rate}
\end{figure}

\subsubsection{Pseudoscalar}\label{sec:pseudo}
\begin{figure}[H]
    \centering
    \includegraphics[width=\textwidth]{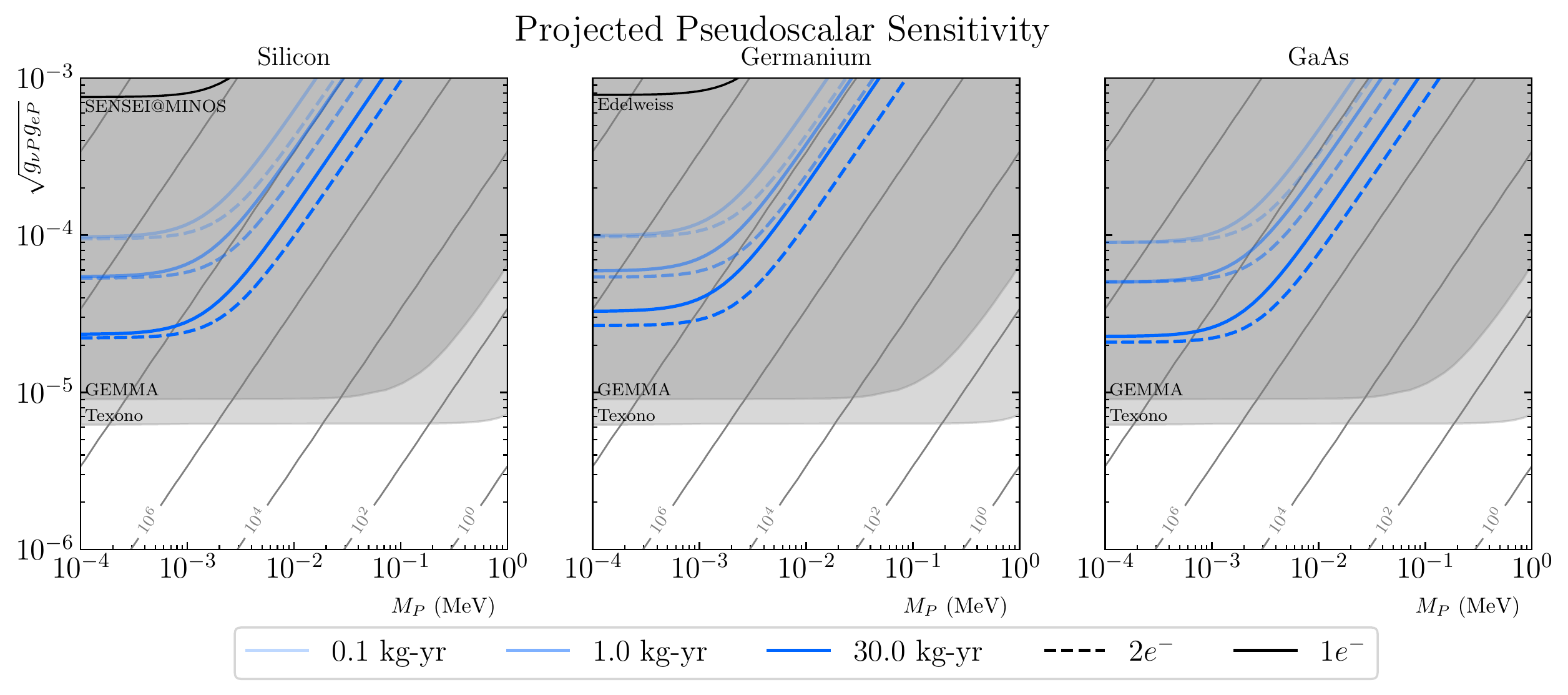}
    \raggedright
    \includegraphics[width=0.81\textwidth]{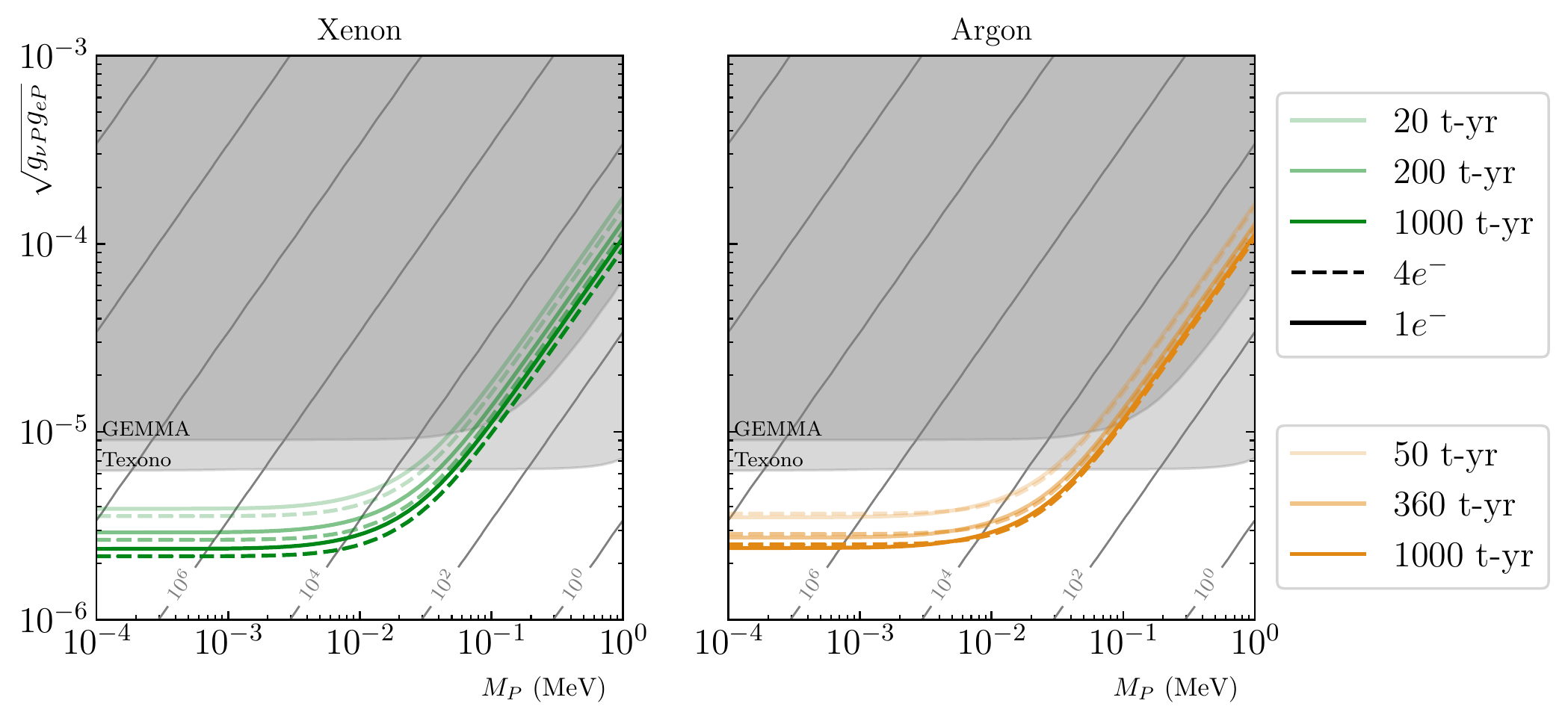}
    \caption{Sensitivity to NSIs mediated by a new pseudoscalar in semiconductor (top) and noble liquid (bottom) targets. The colored lines correspond to the varying exposures, while the solid (dashed) lines correspond to the detector thresholds. Note that, in contrast with the other simplified models, the pseudoscalar only couples to {\bf electrons}. The grey contours represent constant $\epsilon$ as defined in Eq. \eqref{eq:coupling_rel}.}
    \label{fig:pseudo_sens}
\end{figure}

The pseudoscalar mediator case is notably distinct from the other three simplified models in that it does not couple to nuclei, resulting in no enhancement for nuclear scattering (see second line of Eq.~\eqref{eq:n_cross}). Instead, the NSI only appear in the electron-scattering rates, as shown in Fig.~\ref{fig:pseudo_rate}. Therefore, detectors that can distinguish between low-energy nuclear and electron recoils could be a promising way to test such a model.
However, the recoil rate of a light pseudoscalar mediator is not enhanced at low energies, negating the advantages of low-threshold detectors. The top panels of Fig.~\ref{fig:pseudo_sens} show that semiconductor targets require exposures well above the anticipated exposures of future experiments to probe open parameter space. 
The noble liquid detectors are more promising and can probe new parameter space for pseudoscalar masses below $\sim 50$ keV.
\subsection{Magnetic moment}\label{sec:magmoment}
\begin{figure}[h]
    \centering
    \includegraphics[width=\textwidth]{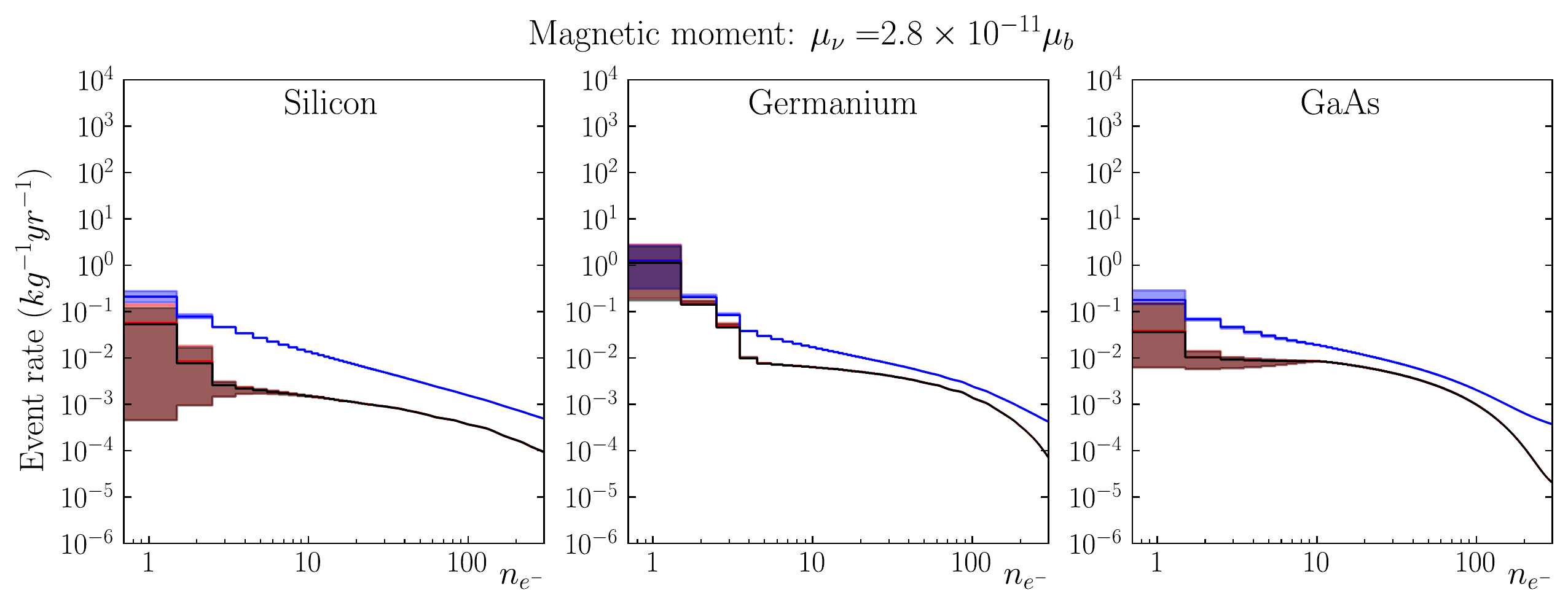}
    \raggedright
    \includegraphics[width=0.91\textwidth]{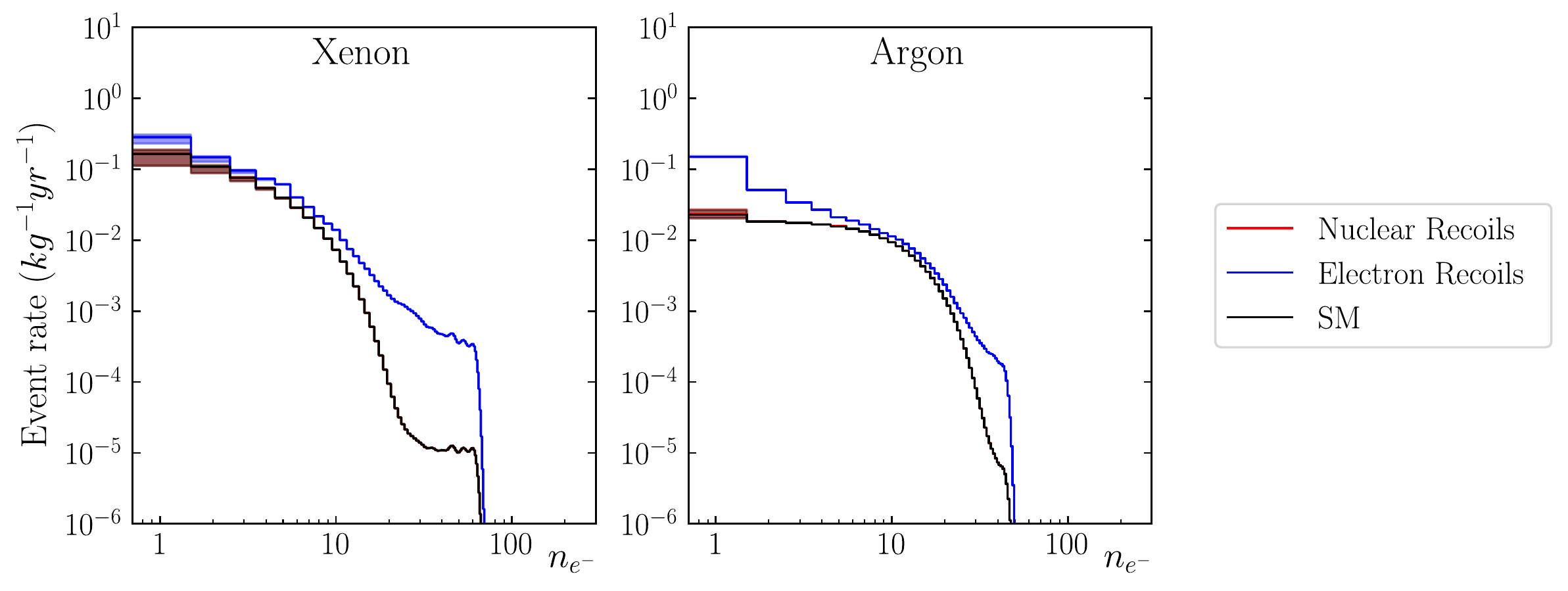}
    \caption{Recoil rates from a neutrino magnetic moment at the Borexino bound. Note that given one free parameter, the nuclear recoil rate (blue) is indistinguishable from the SM rate (black) while the electron rate (red) has a different spectral shape. The uncertainties, denoted by the bands, are dominated by the yield function's effect on the SM nuclear recoils.}
    \label{fig:mag_rates}
\end{figure}

We plot the event rate for the magnetic moment in Fig.~\ref{fig:mag_rates} for a magnetic moment at the Borexino bound.

To predict the sensitivity of a detector to the neutrino magnetic moment, we use a more robust analysis because the model is fully determined by a single parameter. We define the likelihood that a particular distribution of events in an observing run is the result of our signal model (the magnetic moment) or the background (SM),
\begin{equation}\label{likelihood}
    \mathcal{L}(\mu_\nu, \Vec{\phi}) = \frac{e^{-(\eta_\mu+\Sigma_{j=1}^{n_\nu}\eta_j)}}{N!} \times \prod_{j=1}^{n_\nu}\mathcal{L}(\phi_j) \times \prod_{i=1}^{N}\left(\eta_\mu f_\mu(n_i)+\sum_{j=1}^{n_\nu}\eta_j f_j(n_i) \right)\, .
\end{equation}
In so doing, we consider both the total rate of events as well as their distribution in the charge bins. This approach includes the effect of the uncertainties in the solar neutrino fluxes and can be generalized to determine the significance between multiple BSM models.

Our approach follows the analysis in Ref.~\cite{essig_solar_2018} and uses the likelihood function Eq.~\eqref{likelihood} where $\mu_\nu$ and $\Vec{\phi}$ are nuisance parameters corresponding to the magnetic moment and neutrino fluxes respectively. $\eta_\mu$ and $\eta_j$ are the number of events resulting from the magnetic moment and $j$-th component of the solar flux respectively, and $f_\mu(n_i)$ and $f_j(n_i)$ are the (normalized) distributions of events in the $n_i$ bin from the magnetic moment and solar flux components. We then vary the nuisance parameters to maximize the likelihood functions both with $\mu_\nu = 0$ fixed and with $
\mu_\nu$ as a free nuisance parameter. The statistical penalty associated with these variations of the fluxes is captured by
\begin{equation}\label{gauss}
    \mathcal{L}(\phi_j) = \frac{1}{\sigma_j\sqrt{2\pi}} e^{-\frac{1}{2}\left(\frac{\phi_j-\phi_{0,j}}{\sigma_j}\right)^2}\, .
\end{equation}
The ratio of these two minima 

\begin{equation}
    \lambda = \frac{\mathcal{L}_{max}(\mu_\nu=0, \Vec{\phi})}{\mathcal{L}_{max}(\mu_\nu\neq 0,\Vec{\phi})}\, ,
\end{equation}
is used to determine the test statistic: $t = -2\log(\lambda)$ if the magnetic moment which maximizes $\mathcal{L}$ is positive and zero otherwise.

This process gives the significance $\sigma=\sqrt{t}$ of a single data set containing the number of observations in each charge bin. To project a discovery reach, we assume the NSI exists, generate 1000 sets of Poisson fluctuated pseudodata around the rates predicted by the NSI at each exposure or magnetic moment and average the significance.

For the neutrino magnetic moment, we plot the results of our more rigorous analysis and show the projected reach of a 100 kg-yr detector as a function of the magnetic moment and the discovery reach for a fixed magnetic moment $(\mu_\nu = 2.8\times 10^{-11} \mu_b)$ at the Borexino bound as a function of the exposure.

\begin{figure}[h]
\centering
\begin{minipage}{.5\textwidth}
  \centering
  \includegraphics[width=\linewidth]{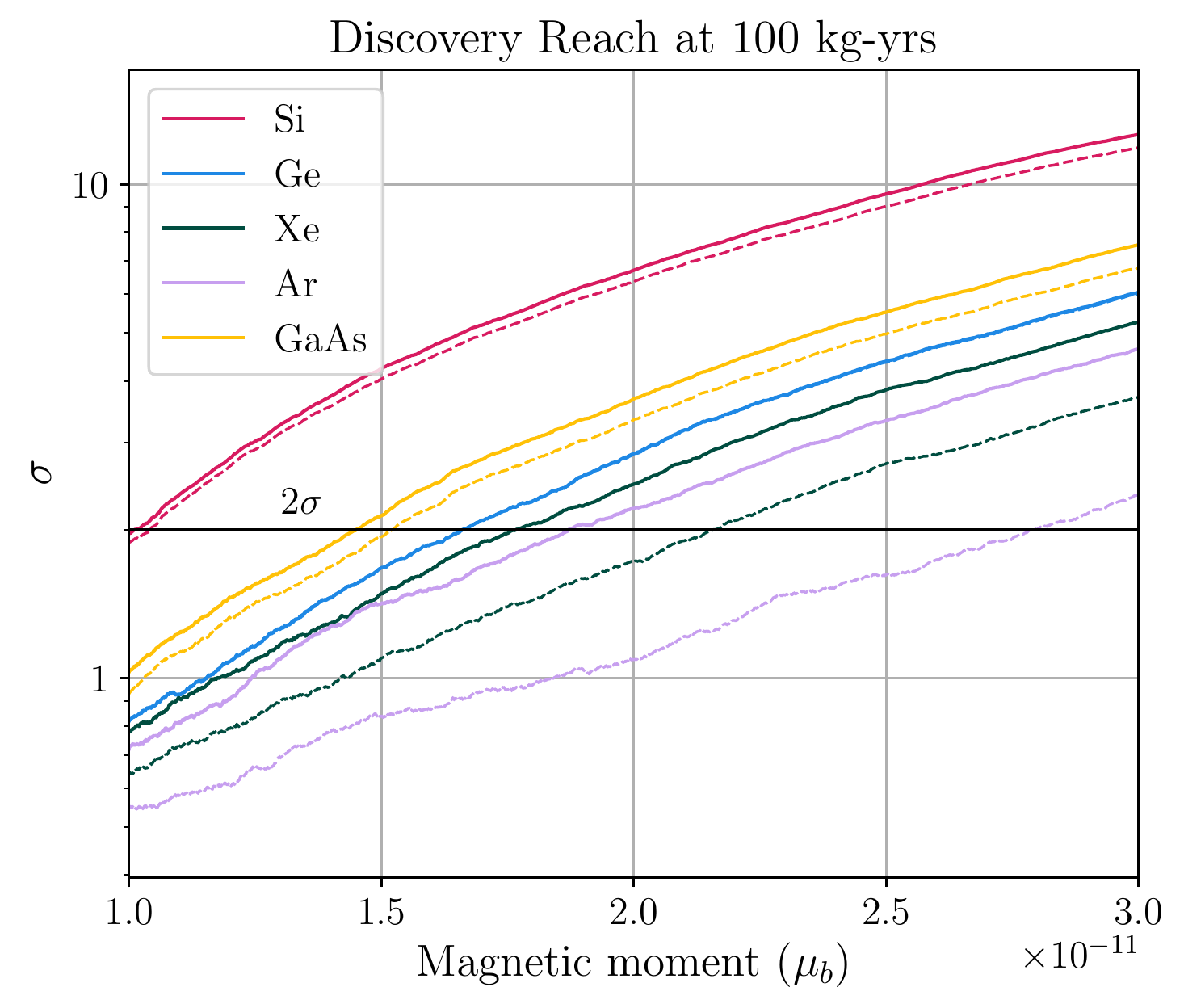}
  \label{fig:sub1}
\end{minipage}%
\begin{minipage}{.5\textwidth}
  \centering
  \includegraphics[width=\linewidth]{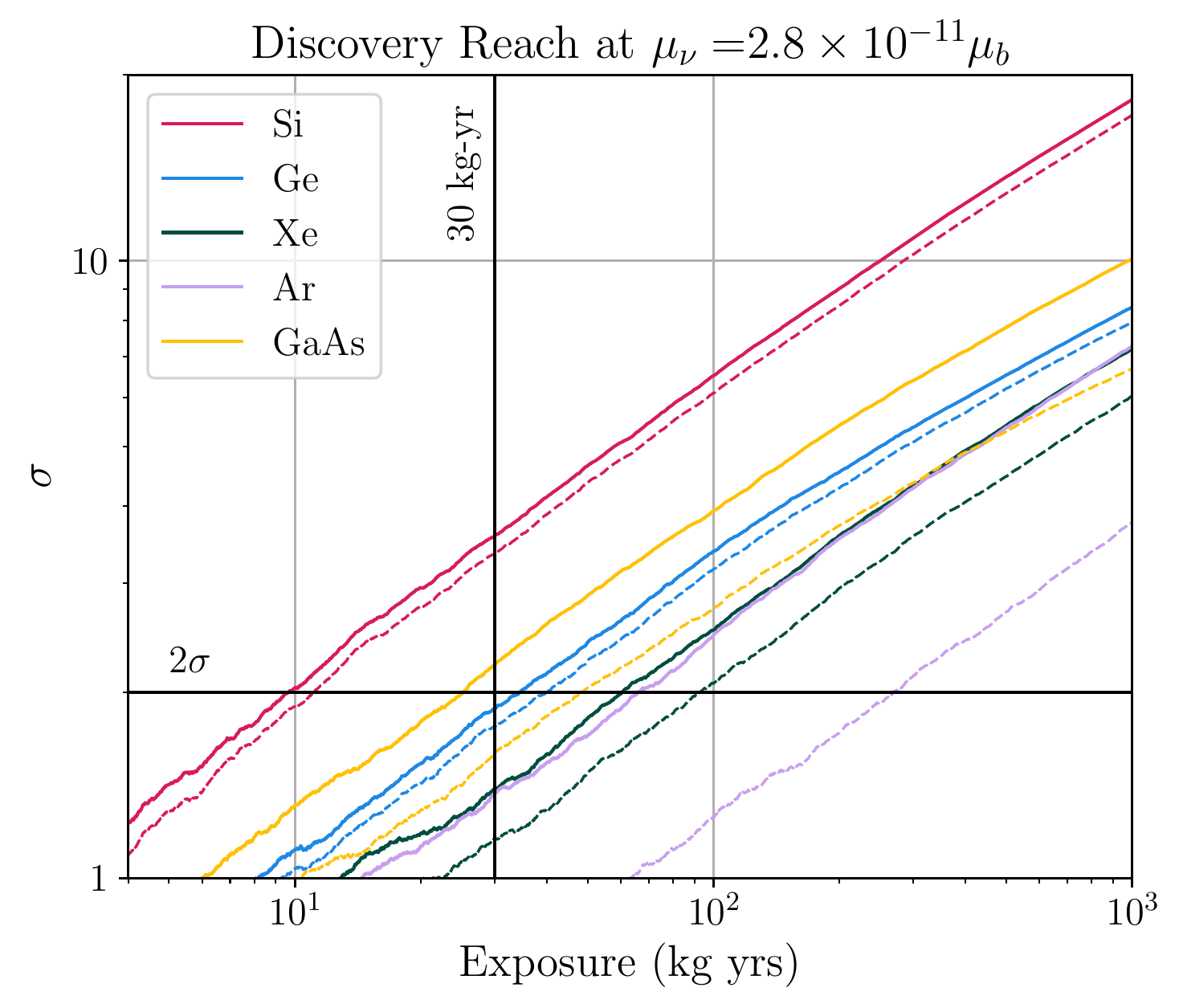}
  \label{fig:sub2}
\end{minipage}
\caption{The discovery reach for the magnetic moment as a function of the value of the magnetic moment (left) and exposure (right). Solid lines are derived using all charge bins, dashed are using $n_e = 2$ in the semiconductors (Si, Ge, GaAs) and $n_e \geq 4$ in Xe and Ar.}
\label{fig:mag_sens}
\end{figure}

\section{Interpreting Simplified Models} \label{sec:discussion}
We have presented our results in the language of simplified models, which focuses on the low-energy Lagrangian without specifying the gauge-invariant models at high-energies. However, one can map our results to motivated  UV-completions to compare the constraints from low-threshold direct-detection experiments and with other model-specific probes. 

\begin{figure}
    \centering
    \includegraphics[width=0.8\textwidth]{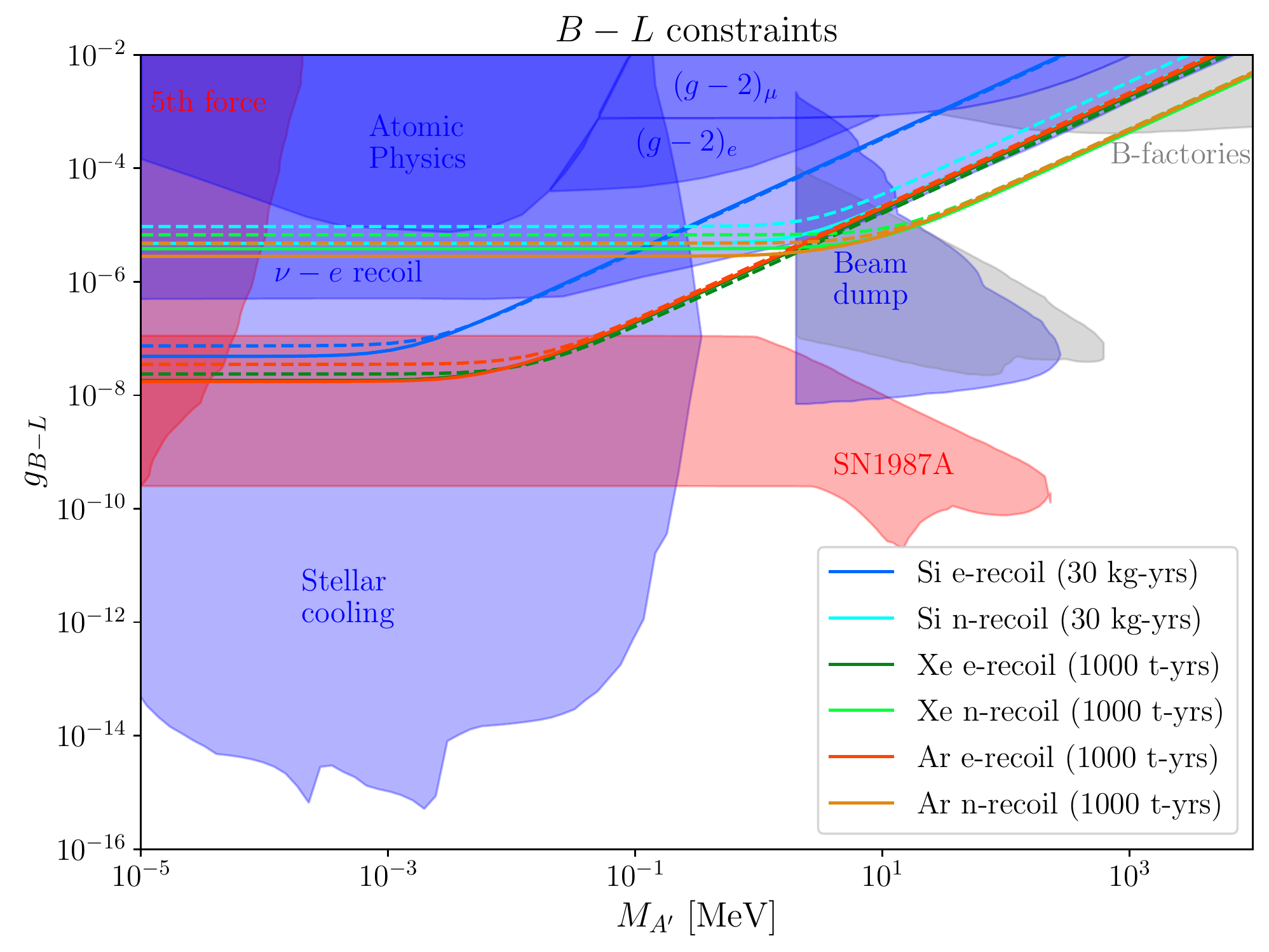}
    \caption{Constraints on gauged $B-L$ from terrestrial and astrophysical sources. Blue regions are sensitive to electron couplings, grey regions require both couplings and red regions are sensitive if the $A'$ couples to either. SN1987A is dominated by neutrino pair coalescence and is therefor relevant regardless of electron and quark couplings~\cite{croon_supernova_2021} as are fifth force searches~\cite{adelberger_fifth_force, BORDAG20011_fifth_force}. The anomalous magnetic moment~\cite{bennett_g-2}, atomic physics~\cite{bartlett_fifth_force}, and stellar cooling~\cite{Redondo_2008, hardy_stellar_2017} constraints require leptonic couplings. A stellar cooling bound can also be produced without an electron coupling by $\nu$ radiation, but this process is subdominant. The beam dump constraints~\cite{bjorken_fixed, batell_fixed} can be divided into two categories: proton and electron beams. Both look for $e^+ e^-$ final states, but the former (grey) requires both electron and baryon couplings while the latter (blue) requires only electron couplings. The B-factories~\cite{cerdeno_physics_2016} constraint relies on both electron and quark couplings. The $\nu - e$ constraint~\cite{cerdeno_physics_2016} uses combined results of GEMMA~\cite{beda_gemma_2013}, Borexino~\cite{Borexino:2008dzn}, Texono~\cite{texono}, and CHARM-II~\cite{Charm_tex}. The blue (cyan) lines represent the projected sensitivity of silicon using electron (nuclear) recoils and line styles the same as Fig.~\ref{fig:vector_semi_sens}. Similarly, green (lime) lines represent the electron (nuclear) detection channels in xenon and dark (light) orange for argon. For noble elements, the line style matches Fig.~\ref{fig:vector_lng_sens}}
    \label{fig:BL}
\end{figure}

As an illustrative example, one can map the vector-mediated model to the well-studied gauged $U(1)_{B-L}$. The Lagrangian for such a model is given by
\begin{equation} \label{BL1}
    \mathcal{L}_{B-L} \supset -g_{B-L}\bar{e}\gamma^\alpha A'_\alpha e +\frac{1}{3}g_{B-L}\bar{q}\gamma^\alpha A'_\alpha q -g_{B-L}\bar{\nu}\gamma^\alpha A'_\alpha \nu - \frac{1}{2}\epsilon F'_{\mu\nu}F^{\mu\nu} \, ,
\end{equation}
where $A'$ is the new vector boson, $g_{B-L}$ as the gauge coupling and $\epsilon$ is the kinetic mixing of the SM photon $A$ and $A'$. 
In addition to the direct-detection constraints presented in this work, the $B-L$ model has constraints from both terrestrial experiments~\cite{bennett_g-2, pospelov_U1, bjorken_fixed, batell_fixed, essig_colliders, essig_discovering, bartlett_fifth_force, Ahlers_LSW, BORDAG20011_fifth_force, adelberger_fifth_force} and astrophysical observables~\cite{croon_supernova_2021, Dent_supernova, Redondo_2008, Arik_2009_cast, Mirizzi_2009_CMB, Fixsen_1996_CMB, Borexino:2008dzn, beda_gemma_2010, hardy_stellar_2017}. In Fig.~\ref{fig:BL}, we show these additional constraints, color coded by whether the constraint depends on couplings to electrons (blue), both electrons and nuclei (grey), or either (red).
A priori, the astrophysical constraints which arise from stellar cooling measurements dominate the parameter space of interest.

Additionally, there is a choice of parameters $\epsilon, g_{B-L}$ such that the gauge contribution exactly cancels with the kinetic mixing of the new vector boson with the photon, which provides an interesting case where the new vector directly couples only to neutrons and neutrinos and couplings to electrons are loop-suppressed.\footnote{However, such a choice presents both a fine-tuning problem and a stability problem under renormalization group evolution (see {\it e.g.} discussion in~\cite{heeck_unbroken_2014}).} This loop-suppression relaxes several of the astrophysical constraints, specifically the ones that rely on the electron-coupling, while preserving the direct-detection constraints. The sensitivities to nuclear couplings plotted in Figs.~\ref{fig:vector_semi_sens} and~\ref{fig:vector_lng_sens} will change modestly in such a model. The cancellation eliminates the couplings to protons which reduces the effective coupling Eq.~\eqref{eq:coherence} by a factor of $A/(A-Z)$, thereby reducing the sensitivity to nuclear scattering. In contrast, the sensitivities to electron-scattering will be greatly reduced as the interactions are cancelled at tree-level and the mediator becomes leptophobic. This alleviates the constraints (shown in blue and gray in Fig.~\ref{fig:BL}) arising from stellar cooling, beam-dump experiments, $\nu-e$ recoils, and $(g-2)$. However, the constraints coming from fifth force experiments~\cite{bartlett_fifth_force, adelberger_fifth_force, BORDAG20011_fifth_force} and SN1987a~\cite{croon_supernova_2021, Dent_supernova, Redondo_2008} still apply as these rely on the coupling to nucleons. Chameleon effects are another proposed way to evade or weaken astrophysical constraints~\cite{Feldman_2006}. 

In addition to $U(1)_{B-L}$, a vector model can be mapped to $U(1)_{L_\mu - L_\tau}$ or the kinetic mixing of heavy sterile neutrinos. The latter can be tuned in such a way that the heavy neutrinos only produce recoils at arbitrarily low energy to evade constraints from existing detectors even with a much larger coupling. Low-threshold detectors provide an excellent way to study such a model.

Constraining scalar-mediated interactions of neutrinos is relevant as such models can influence neutrino oscillations in matter by modifying the effective mass matrix~\cite{babu_neutrino_2020}. This effect is independent of the neutrino energy, unlike the vector or axial vector cases which modify the matter potential. Since the scalar interactions change the effective mass matrix, they change the CP phase and can lead to fake CP violation~\cite{Shao-Feng_scalars}.

Axial vector currents would evade the next generation of long baseline neutrino oscillation experiments which will be able to place strong constraints on vector currents due to their influence on the matter potential~\cite{Miranda_2015}. This makes the improvement seen in Figs.~\ref{fig:axial_semi_sens} and \ref{fig:axial_lng_sens} more important.

Note that we have not provided an exhaustive list of UV complete theories, as the set of simplified NSI models discussed in this paper are common to a large set of theories. 

\section{Conclusions} \label{sec:conclusions}

In this work, we have shown that near-future low-threshold dark matter detectors, both semiconductor and noble liquid targets, have the potential to probe previously unconstrained parameter space for vector, scalar, and axial vector NSI simplified models; pseudoscalar models are more difficult to probe.
The potential for vector mediators is especially promising with improvements to the electron (nuclear) coupling constraints of an order of magnitude for $M_V\lesssim 1 (1000)$ keV in semiconductors and one (two) order of magnitude in noble liquid detectors at similar mediator masses. Axial vector models see a similar improvement except where detector nuclei have little or no spin (such as argon). In scalar models, semiconductors can roughly match the electron coupling constraints while improving upon the nuclear ones by an order of magnitude below $M_S\lesssim 1 (1000)$ keV and noble liquids can improve constraints by an order of magnitude for electron and nuclear couplings. In contrast, there is little room for improvement with the pseudoscalar models; the noble liquid detectors can improve upon current constraints by a factor of a few but the semiconductor experiments are not competitive. In the event of a putative signal, one can perform an model-specific analysis in lines with the one discussed in Sec.~\ref{sec:magmoment}.
In addition to the simplified models, both semiconductors and noble liquids will be sensitive to a neutrino magnetic moment. A silicon (germanium) detector can match the current best bounds at 95\%CL with an exposure of 10 (35) kg-yrs while xenon (argon) can do so with 60 (70) kg-yrs. 

The sensitivity of Skipper-CCDs to reactor neutrinos was recently studied in \cite{fernandez-moroni_physics_2021}, which provides a complementary analysis to the one performed in this work. Specifically, there are two key differences between reactor and solar neutrinos: the neutrino energies and the neutrino flavors probed. Although solar neutrinos are a lower intensity source, they are at higher energy than their reactor counterparts. The other difference is in the neutrino flavors detected. For a detector $\sim 10$m from a reactor, all of the incident neutrinos will be electron neutrinos. In comparison, solar neutrinos are traveling much longer distances so a detector using solar neutrinos will be sensitive to the effects of neutrino oscillations and NSIs with any neutrino flavor. This broadens the set of UV theories such as $U(1)_{L_\mu - L_\tau}$ which map to these simplified models. Combining the results from reactor and solar neutrinos broadens the scope of analyses on NSIs.  
 
In summary, we have demonstrated that current and near-future low-threshold DM direct detection experiments have tremendous potential to probe models of NSIs. The semiconductor experiments can improve the existing constraints on the neutrino magnetic moment and to measure the changes to the SM CEvNS rate while liquid xenon and argon detectors will be able to improve the measurement of potential electron couplings. We have shown that in noble liquid detectors, the improvements gained by lowering the threshold have a stronger effect at low masses compared to the order of magnitude increase in the exposure. A more robust statistical analysis could generally improve these constraints, but we leave that analysis to be performed on real data collected by future detectors. We have omitted the uncertainties associated with our choice of yield function to maintain plot legibility, but we include them in Appendix~\ref{app:yield} for a single exposure to demonstrate their importance. They dominate the uncertainty in these calculations and better measurement of them, specifically at low energies, is vital to validating constraints placed by such low-threshold DM detectors.

We have made use of the existing infrastructure of DM detectors to project constraints on NSIs because the signals produced by these NSIs, specifically in nuclear recoils, are very similar to DM recoils. A single event is indistinguishable, but by using the distribution of events a robust likelihood analysis, like the one used for the neutrino magnetic moment, can disentangle two BSM models such as a NSI and an ultralight DM particle. Additional information can be seen by the relative strengths of a signal in electron recoils or different materials such as argon's insensitivity to axial vector models. Such dedicated analyses are worthwhile undertakings that we leave for future work.
\section*{Acknowledgments}
We thank Haidar Esseili, Guillermo Fernandez-Moroni, and Youssef Sarkis Mobarak for many useful discussions, as well as Roni Harnik, Joachim Kopp, and Pedro Machado for comments on the manuscript. This work was supported in part by NSF CAREER grant PHY-1944826. 

\bibliographystyle{utphys}
\bibliography{NSI_detection.bib}

\appendix
\section{Yield Function Uncertainties} \label{app:yield}
As noted in Sec.~\ref{sec:discussion}, the dominant source of uncertainty in these projected sensitivities in semiconductors (or the actual constraints placed by an experiment) is the yield function. Here we demonstrate these effects on the nuclear recoil sensitivities at a single exposure (1 kg-yr in semiconductors, 100 t-yrs in Xe and Ar) for both the 1 and 2$e^-$ thresholds in semiconductors and 1 and 4$e^-$ thresholds in the noble liquids. The sensitivity to electron recoils is only influenced by the change in the SM CEvNS rate thus is less sensitive to the yield function compared to the nuclear recoil case.

\begin{figure}[H]
\centering
\begin{minipage}{.425\textwidth}
  \centering
  \includegraphics[width=\linewidth]{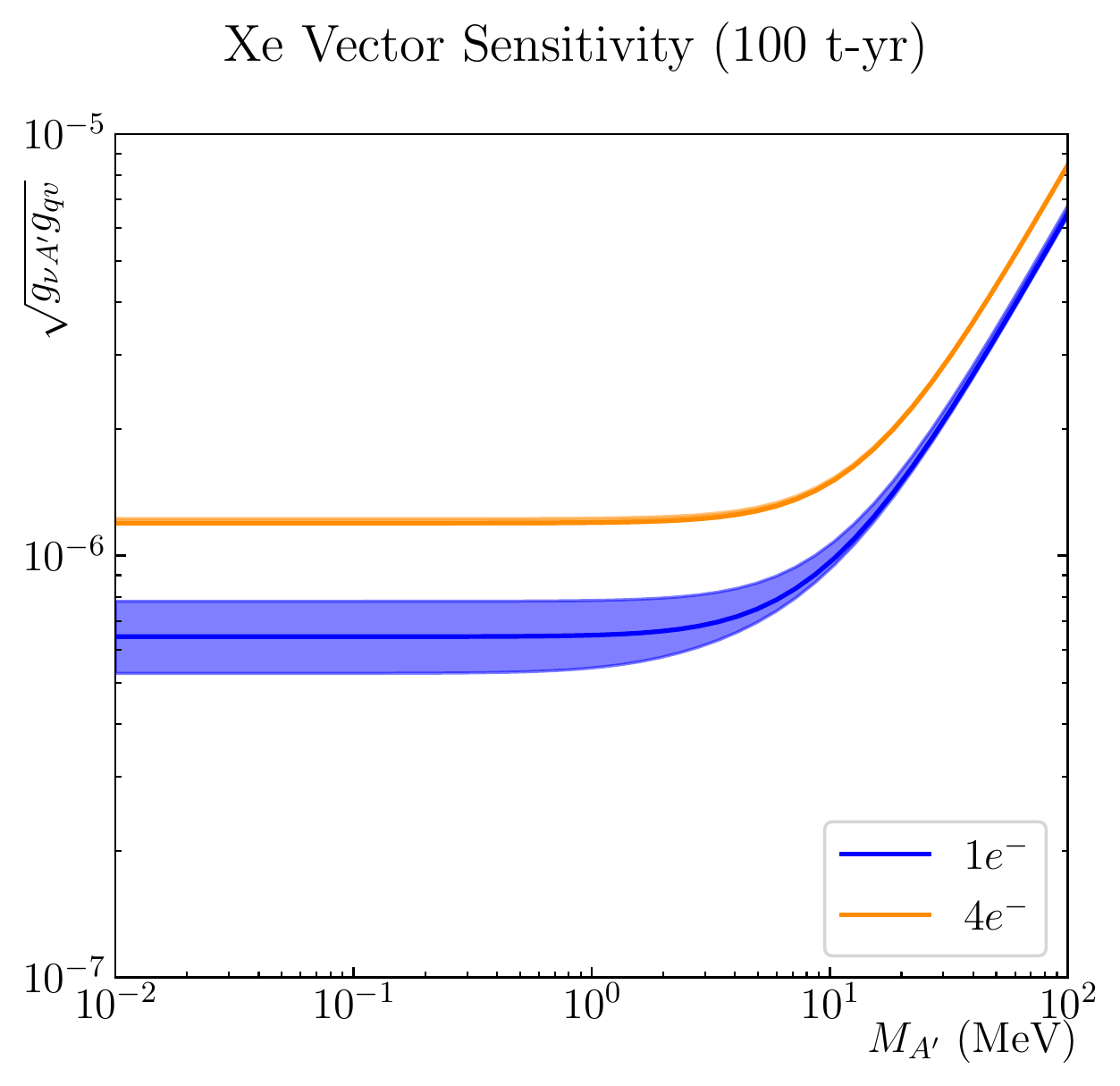}
  \includegraphics[width=\linewidth]{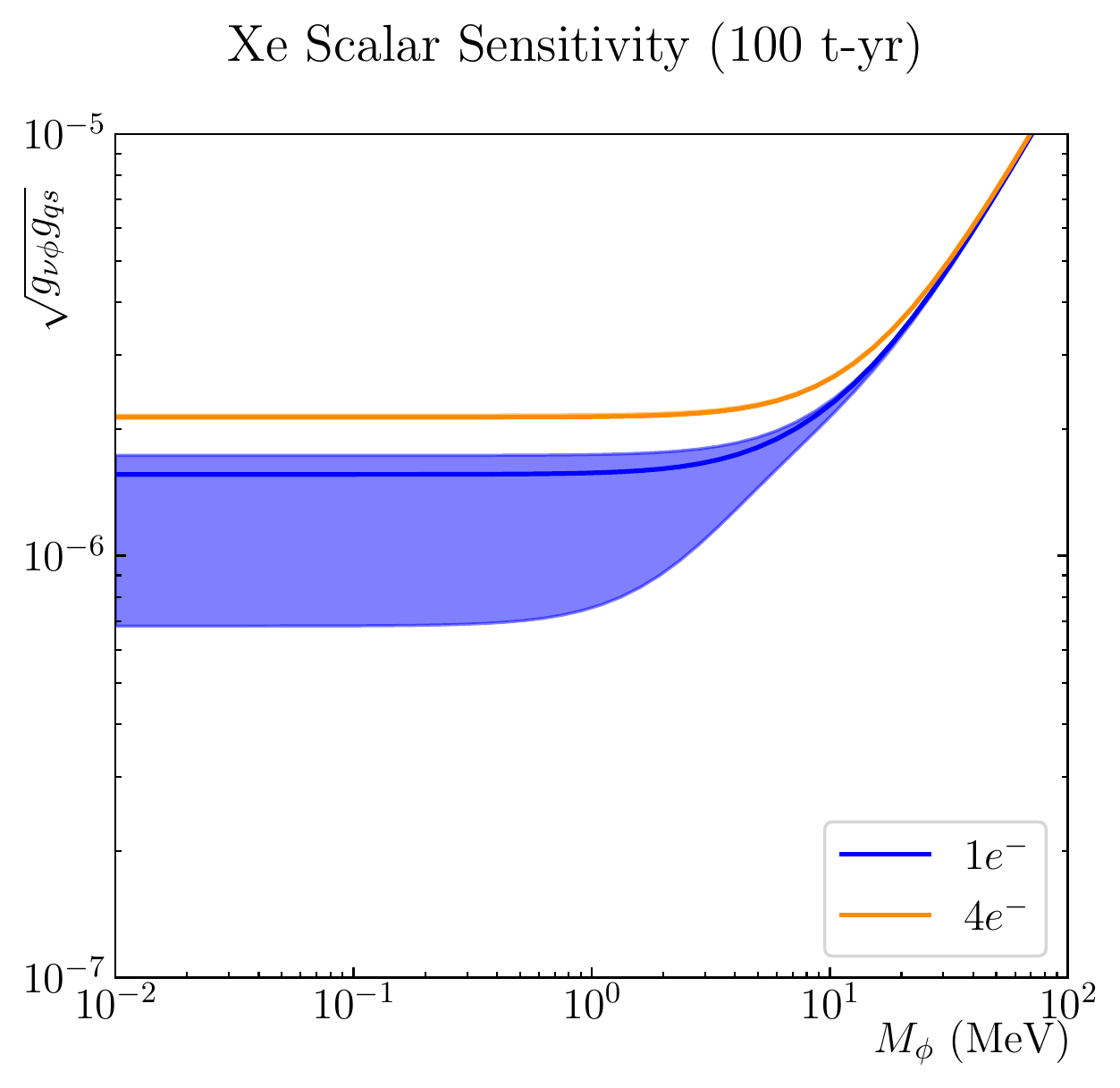}
\end{minipage}
\begin{minipage}{.425\textwidth}
  \includegraphics[width=\linewidth]{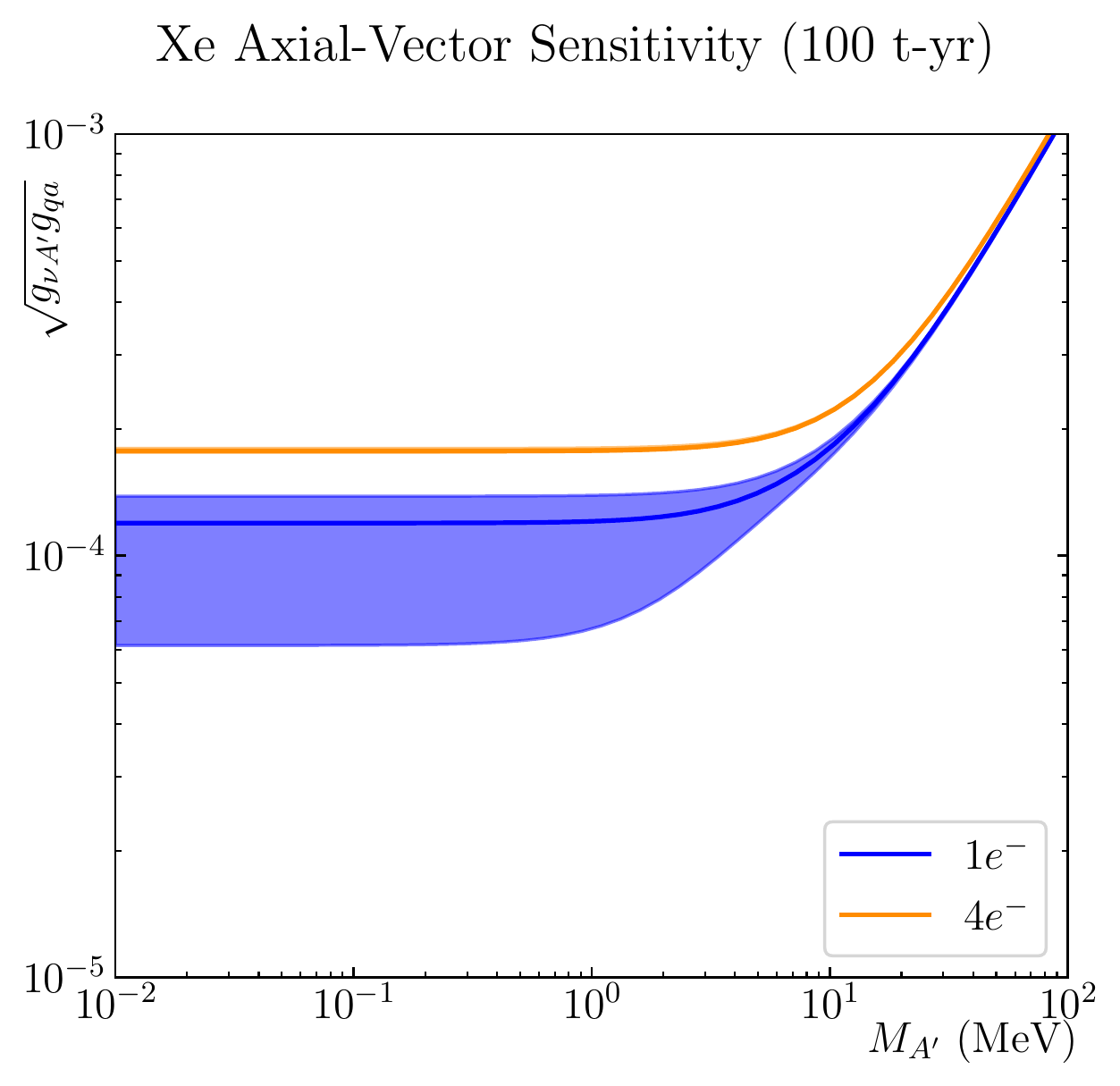}
  \vspace{16mm}
  \caption{In xenon and argon (not shown) the $1e^-$ threshold sensitivity is sensitive to the yield function, but the $4e^-$ case is not since $4e^-$ and higher events correspond to energies where the yield functions are well measured.}
  \label{fig:lng_unc}
  \vspace{16mm}
\end{minipage}
\end{figure}

\begin{figure}
\centering
\begin{minipage}{.5\textwidth}
  \centering
  \includegraphics[width=0.85\linewidth]{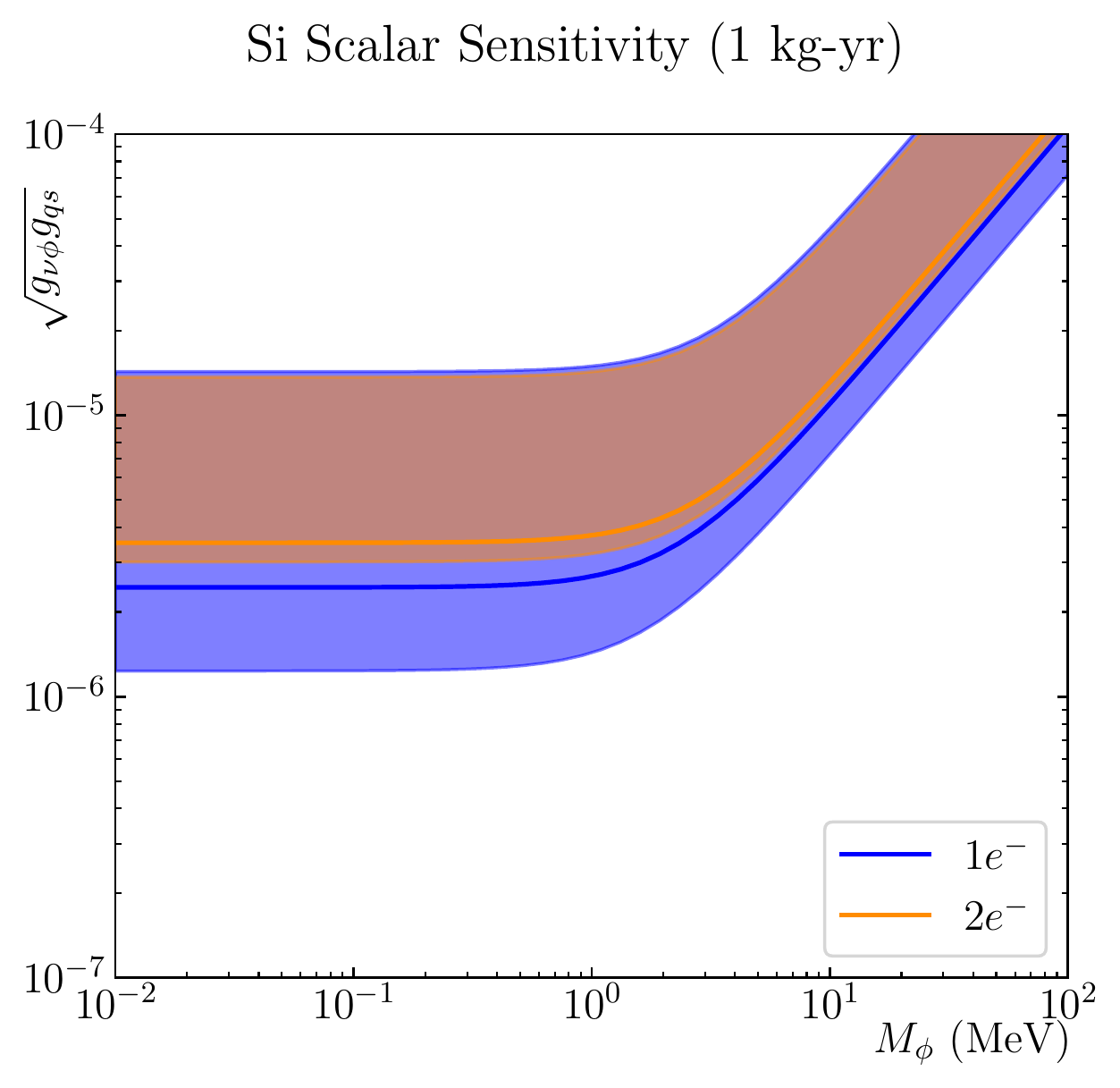}
  \includegraphics[width=0.85\linewidth]{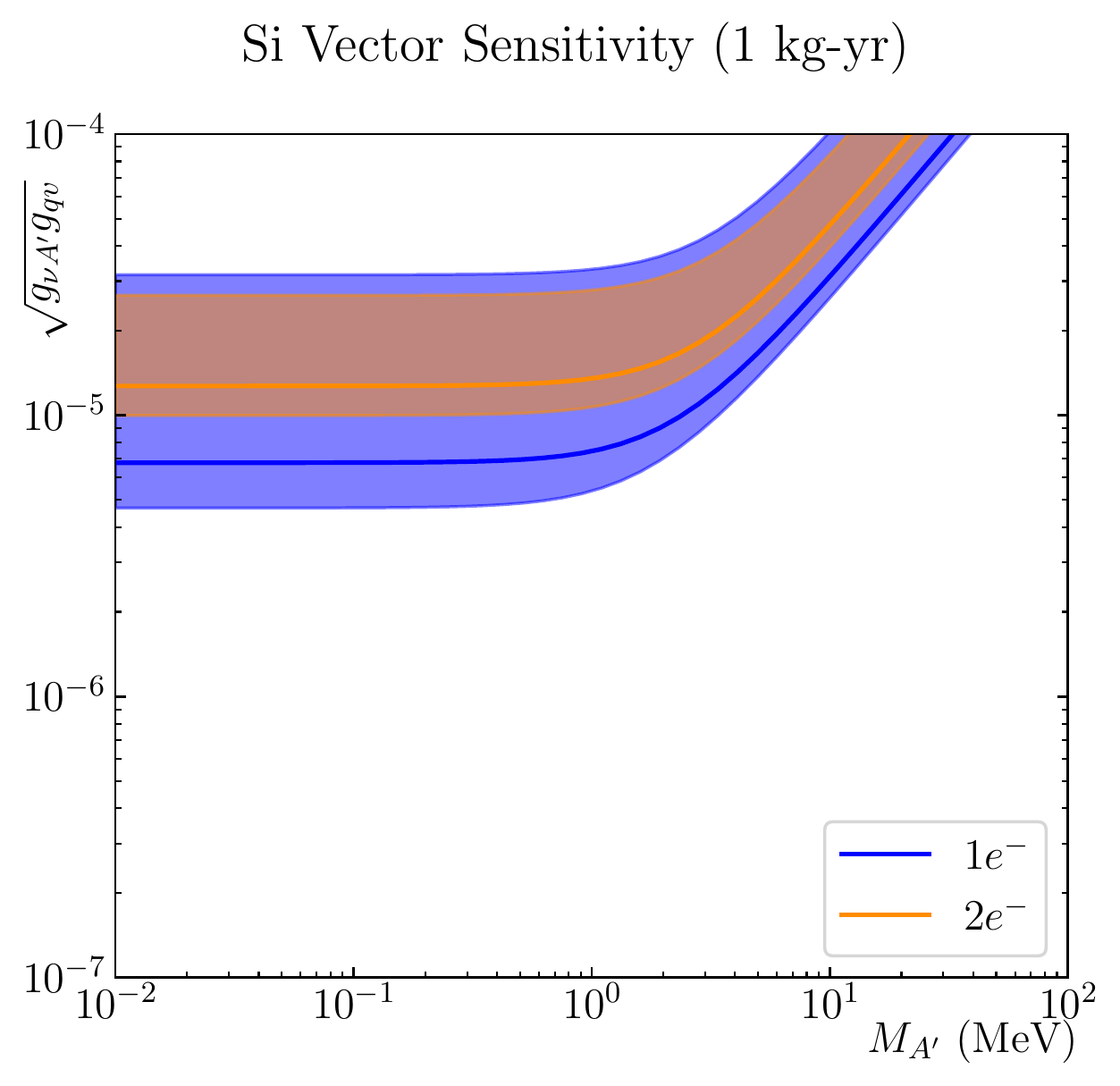}
  \includegraphics[width=0.85\linewidth]{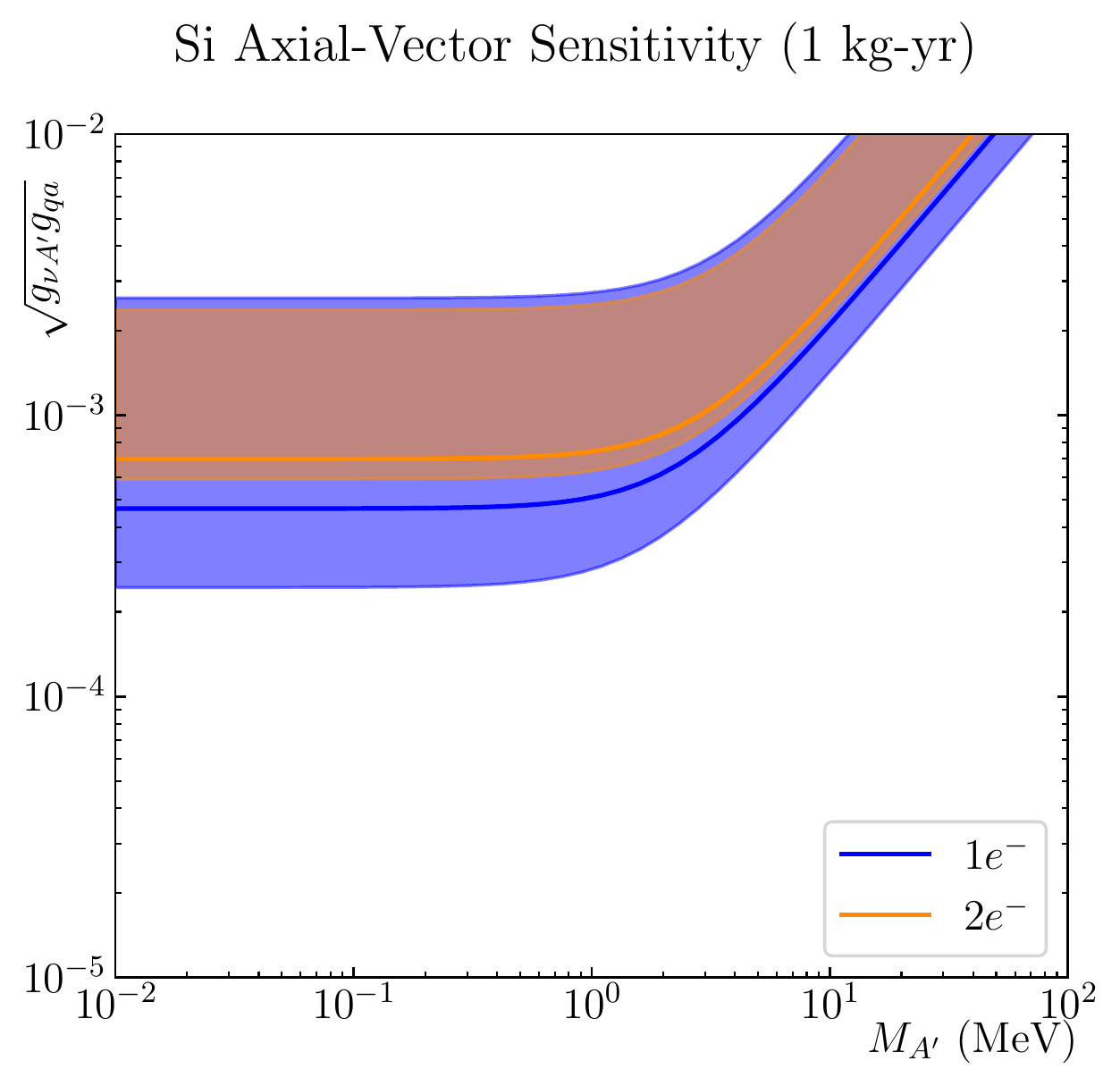}
\end{minipage}%
\begin{minipage}{.5\textwidth}
  \centering
  \includegraphics[width=0.85\linewidth]{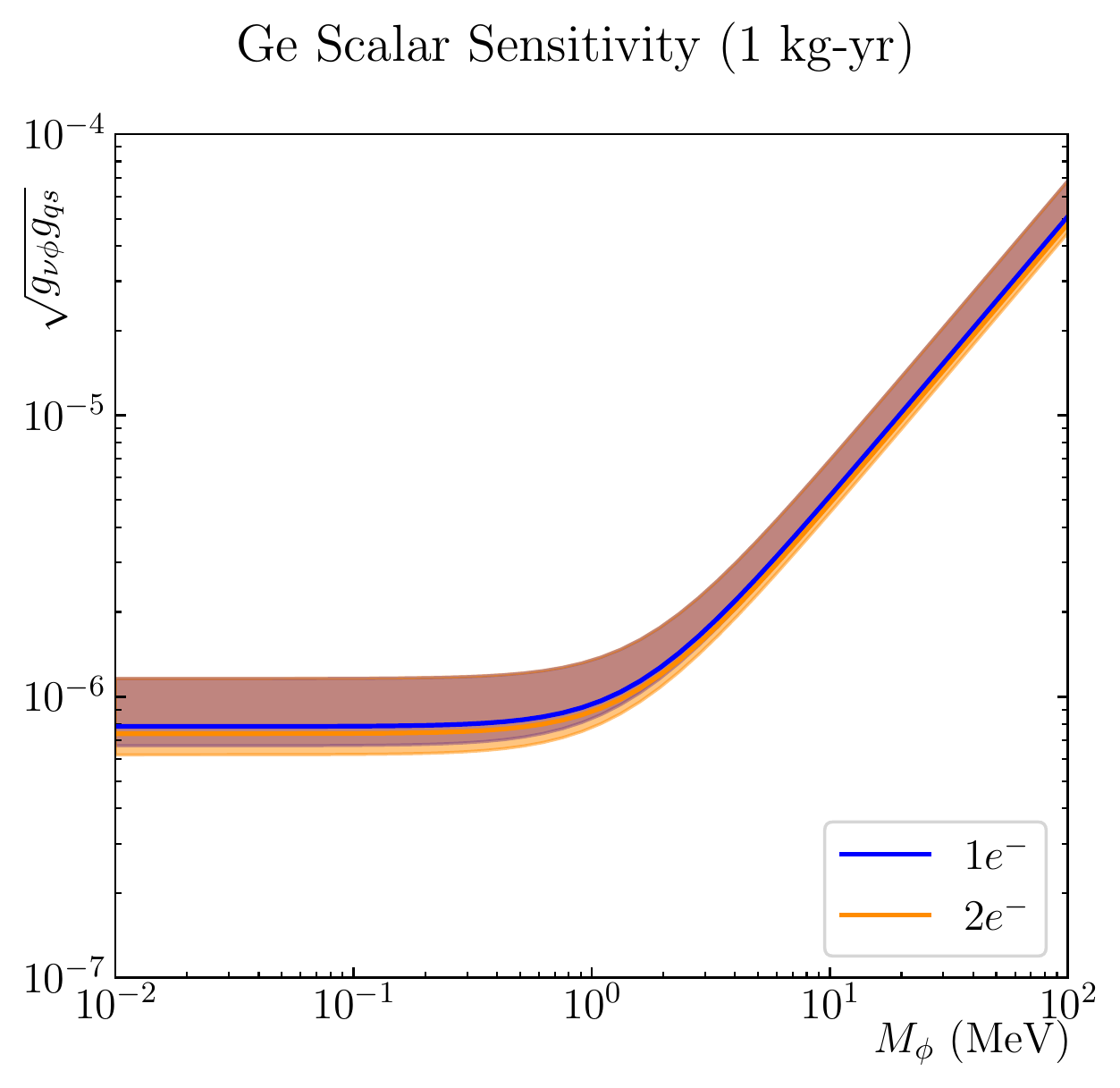}
  \includegraphics[width=0.85\linewidth]{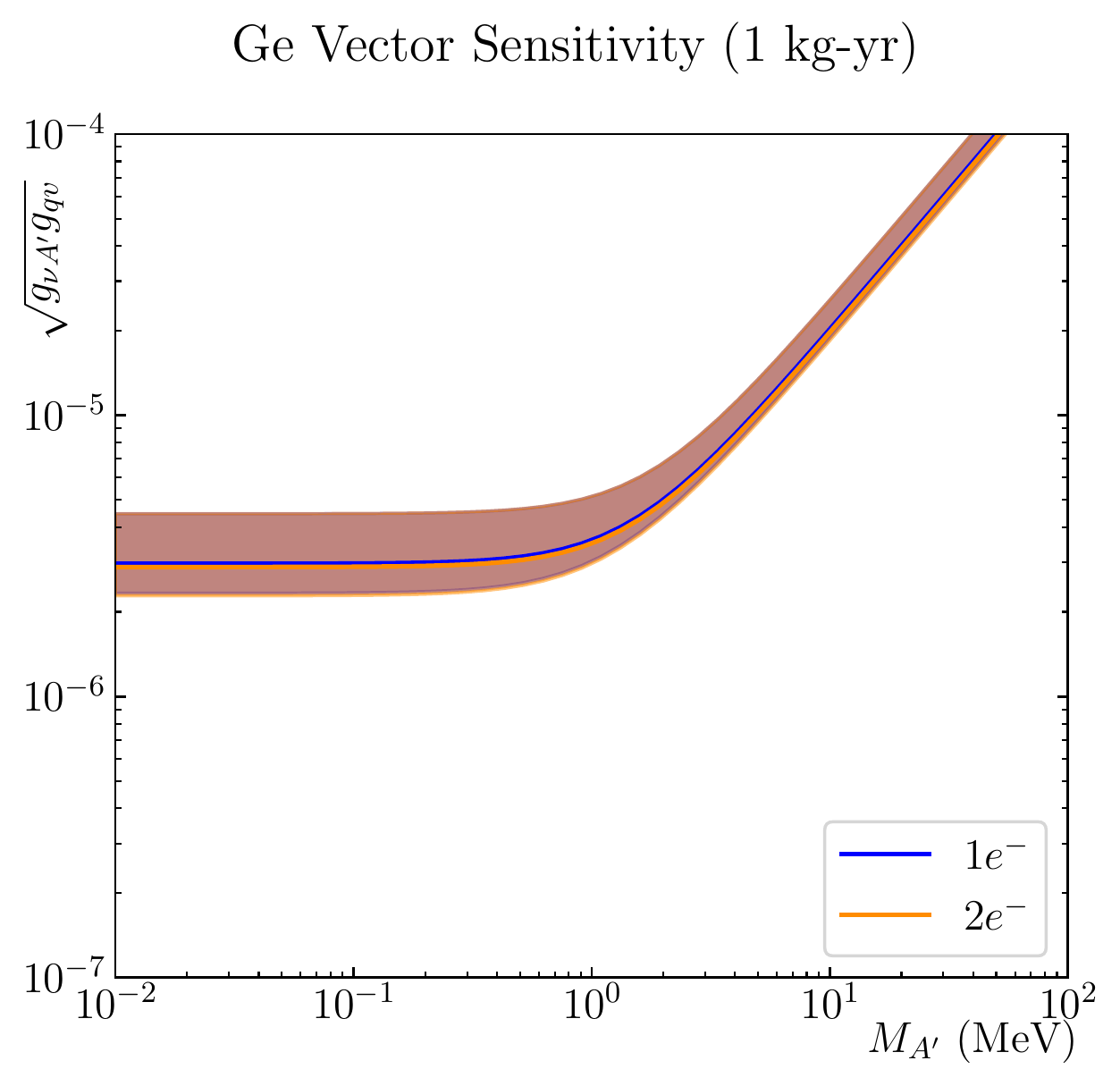}
  \includegraphics[width=0.85\linewidth]{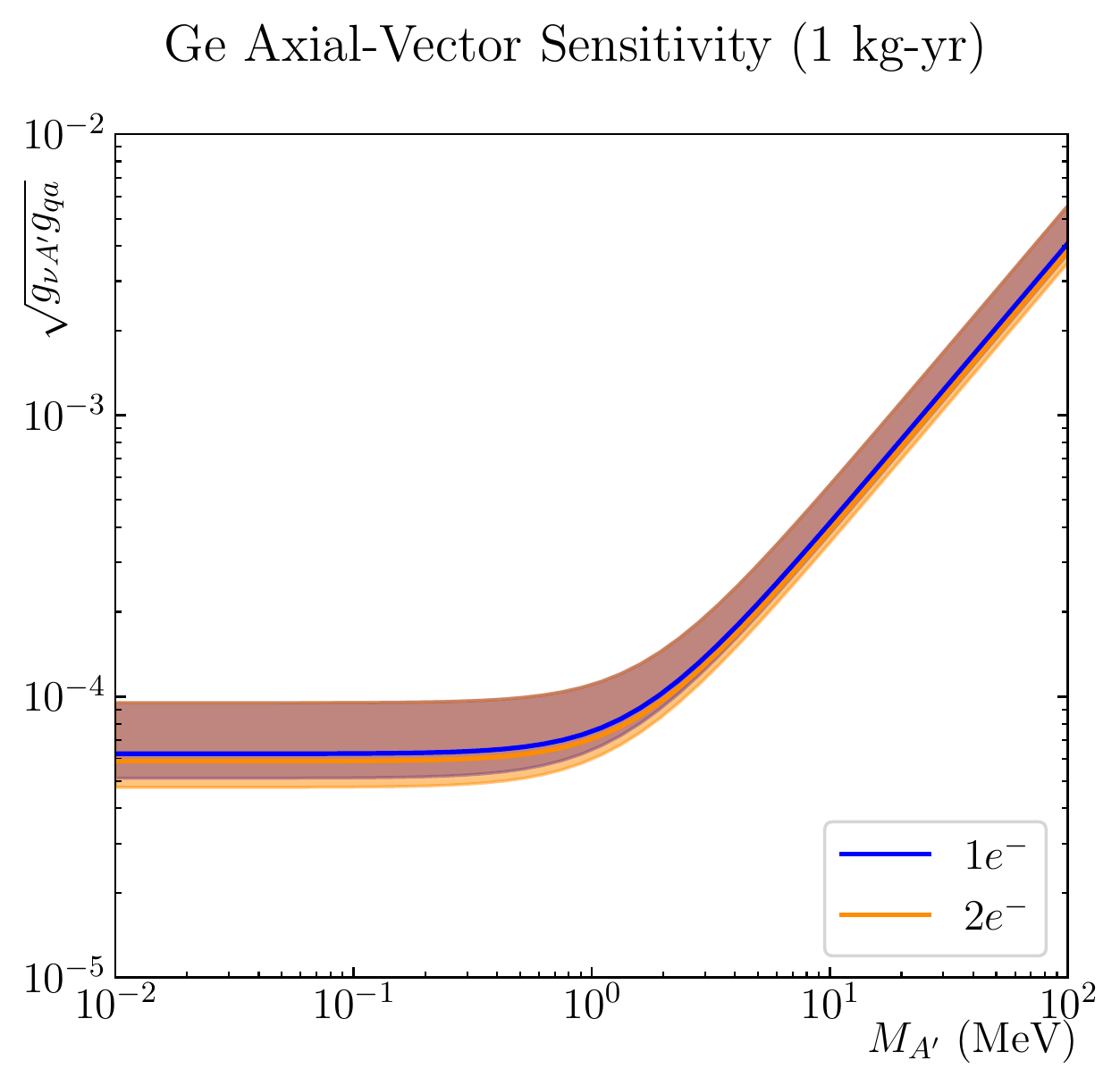}
\end{minipage}
\caption{Colored bands indicate the range of possible sensitivities to nuclear recoils contained within the choice of yield function between the red and blue lines in Fig.~\ref{fig:yield} for a 1 kg-yr semiconductor. Colors indicate the detector threshold.}
\label{fig:semi-cond_unc}
\end{figure}

In Fig.~\ref{fig:lng_unc} we plot only the xenon rates, noting that argon exhibits a similar dependence on the yield function except in the axial vector case to which it is not sensitive. In Fig.~\ref{fig:semi-cond_unc}, we plot the uncertainty in silicon and germanium. We see that the ionization yield uncertainties have stronger effect for the lower threshold (1$e^-$), where they can change the projections by up to an order of magnitude in silicon. 

\end{document}